\documentclass[prd,reprint,superscriptaddress,preprintnumbers,nofootinbib,amsmath,amssymb,aps,onecolumn]{revtex4-2}

\usepackage{graphicx}
\usepackage[dvipsnames,HTML]{xcolor}
\usepackage[colorlinks=true,citecolor=blue,urlcolor=blue]{hyperref}
\definecolor{myred}{HTML}{D81B60}
\definecolor{myblue}{HTML}{1E88E5}
\definecolor{myyellow}{HTML}{FFC107}
\definecolor{mygreen}{HTML}{004D40}

\begin{document}
\preprint{UT-WI-39-2025}

\title{Early Universe Constraints on Variations in Fundamental Constants Induced by Ultralight Scalar Dark Matter}

\author{Subhajit Ghosh}
\affiliation{Texas Center for Cosmology and Astroparticle Physics, Weinberg Institute, Department of Physics, The University of Texas at Austin, Austin, TX  78712, USA}

\author{Kimberly K.~Boddy}
\affiliation{Texas Center for Cosmology and Astroparticle Physics, Weinberg Institute, Department of Physics, The University of Texas at Austin, Austin, TX  78712, USA}

\author{Tien-Tien Yu}
\affiliation{Institute for Fundamental Science and Department of Physics, University of Oregon, Eugene, OR 97403, USA}

\begin{abstract}
We study the cosmological impact of ultralight dark matter (ULDM) with a quadratic coupling to Standard Model particles. In addition to the suppression of small-scale power from ULDM itself, the coupling induces a variation of fundamental constants that is modulated by the ULDM oscillatory field value. In this work, we consider the ULDM-induced, time-dependent variation of the fine structure constant and the mass of the electron. These variations modify the predicted abundance of light elements during Big Bang nucleosynthesis (BBN) and the process of recombination, thereby affecting the anisotropies of the cosmic microwave background (CMB). We use CMB anisotropy data and baryon acoustic oscillation measurements to obtain constraints on the variation of couplings over a wide range of ULDM masses. We self-consistently account for the modification of the primordial helium abundance during BBN in computing the CMB power spectra. We find that the allowed ULDM fraction of total dark matter abundance is more constrained for ULDM masses $\lesssim 10^{-26}~\mathrm{eV}$ in the presence of the variations. Moreover, our constraints on the variational couplings for ULDM masses $\lesssim 10^{-27}~\mathrm{eV}$ are stronger than the ones derived from the primordial helium abundance at BBN. Under our ULDM model, the variation of fundamental constants has no appreciable impact on the Hubble constant inferred from CMB data and thus does not present a viable solution to the Hubble tension.
\end{abstract}

\maketitle

%%%%%%%%%%%%%%%%%%%%%%%%%%%%%%%%%%%%%%%%%%%%%%%%%%%%%%%%%%%%%%%%%%%%%%%%%%%%%%%
\section{Introduction}

Measurements of the fundamental constants of nature have reached exquisite precision, especially for constants related to atomic physics, such as the fine structure constant and the masses of the proton and electron~\cite{ParticleDataGroup:2024cfk,Mohr:2024kco}.
The possible variation of fundamental constants (VFC) is strongly constrained by terrestrial experiments and by cosmological and astrophysical observations~\cite{Martins:2017yxk,Uzan:2024ded}.
Early Universe studies have constrained VFCs through their modifications of the reaction rates that set the light-element abundances during Big Bang nucleosynthesis (BBN)~\cite{Kolb:1985sj,Campbell:1994bf,Olive:2002tz,Coc:2006sx,Muller:2004gu,Dent:2007zu,Berengut:2009js,Alvey:2019ctk,Martins:2020syb,Sibiryakov:2020eir,Bouley:2022eer,Meissner:2023voo} and modifications to helium and hydrogen recombination during the formation of the cosmic microwave background (CMB)~\cite{Kaplinghat:1998ry,Hannestad:1998xp,Battye:2000ds,Avelino:2000ea,Avelino:2001nr,Martins:2003pe,scóccola2009wmap5yearconstraintstime,Martins:2010gu,Menegoni:2012tq,Planck:2014ylh,Hart:2017ndk,Hart:2019dxi,Sekiguchi:2020teg,Hart:2022agu,Sekiguchi:2020teg,Tohfa:2023zip,Seto:2022xgx,Baryakhtar:2024rky,Baryakhtar:2025uxs,Schoneberg:2024ynd}.

Many BBN studies analyzed VFCs under well-motivated physical models, such as theories with scalar fields~\cite{Bekenstein:1982eu,Damour:1994zq,Damour:2002nv,Olive:2002tz,Chiba:2006xx,Damour:2010rp,Stadnik:2015kia,Bouley:2022eer,Sibiryakov:2020eir,Barrow:2005qf}, extra dimensional models~\cite{Kolb:1985sj,Loren-Aguilar:2003qtx,Li:2005aia,Steinhardt:2010ij}, phase transitions~\cite{Chacko:2002mf,Anchordoqui:2003ij}, and grand unified theories~\cite{Campbell:1994bf,Coc:2006sx,Calmet:2006sc,Dine:2002ir,Calmet:2001nu,Langacker:2001td}.
However, most CMB studies of VFC considered simplified parameterizations---such as a constant shift in the value of the constants or a variation that scales as a power law in redshift---and focused on modifying the recombination era only, though some have considered BBN as well~\cite{Seto:2022xgx,Schoneberg:2024ynd,Baryakhtar:2024rky,Baryakhtar:2025uxs}.
Constant parameterizations, in particular, can be problematic in connecting to models, since there must be a mechanism to relax the variation at late times in order to be consistent with experimental measurements. Even when physical models are considered in CMB analyses~\cite{Zhang:2022ujw}, the impact on BBN is not incorporated or accounted for self-consistently in the analysis.
Additionally, redshift-independent variations of the electron mass---and to a lesser extent, the fine structure constant---during (only) the CMB era has been proffered as a possible mechanism to help alleviate the Hubble tension by increasing the inferred value of the Hubble constant $H_0$~\cite{Hart:2019dxi,Sekiguchi:2020teg,toda_constraints_2025,Hart:2021kad,Seto:2022xgx,Lee:2022gzh,Schoneberg:2021qvd,Zhang:2022ujw,Schoneberg:2024ynd,Baryakhtar:2024rky,Baryakhtar:2025uxs,ACT:2025tim}.
A self-consistent study between at least BBN and the CMB is critical for assessing the viability of VFCs to resolve the Hubble tension.

A well-motivated model of VFCs arises from an ultralight scalar (or pseudoscalar) field $\phi$ coupled to the Standard Model (SM) Lagrangian in a dilaton-like fashion~\cite{Damour:1994zq,Cho:1998js,Damour:2010rp,Damour:2010rm,Cho:2007cy,Hees:2018fpg,Stadnik:2015kia,Stadnik:2015xbn}.
The presence of such a field itself has an impact on the CMB that depends on the form of its potential.
At early times, the field is frozen on its potential at its misaligned initial field value due to Hubble friction, and it contributes as a dark energy (DE) component of the Universe.
At later times, the field oscillates in its potential around its presumed global minimum. For a quadratic potential, the field behaves as a matter component of the Universe, thereby constituting a portion of the total dark matter energy density.
Ultralight dark matter (ULDM) in the form of a scalar, axion, or axion-like field has been studied extensively in the context of the CMB~\cite{Hu:1998kj,Hu:2000ke,Amendola:2005ad,Duffy:2009ig,Arvanitaki:2009fg,Hlozek:2014lca,Marsh:2010wq,Marsh:2015xka,Hlozek:2017zzf,Rogers:2023ezo,ACT:2025tim}.
Moreover, density fluctuations of the ULDM field are suppressed at scales below its coherence length of oscillations~\cite{Marsh:2010wq} and are thus constrained by small-scale structure formation~\cite{DES:2020fxi,Rogers:2020ltq,Park:2022lel,Rogers:2023upm,Nadler:2025fcv,lazare_constraints_2025}.

In this work, we consider the cosmological impact of a subcomponent of dark matter in the form of ULDM with a potential $V(\phi) = m_\phi^2 \phi^2 / 2$, where $m_\phi$ is the mass of the ULDM field $\phi$.
Additionally, the ULDM field is quadratically coupled to the SM, inducing VFCs proportional to $\phi^2$.
We use \textit{Planck} 2018 and SPT-3G CMB anisotropy measurements, as well as baryon acoustic oscillation (BAO) data from SDSS and BOSS, to set self-consistent cosmological constraints on ULDM with VFCs.
To make connections with prior CMB work, we focus on couplings that induce variations in the electron mass and the fine structure constant.
We account for the background evolution and density perturbations of the ULDM field, as well as the $\sim \phi^2$ shift in the electron mass and fine structure constant throughout the BBN and CMB eras.
We capture these effects in our Boltzmann code \texttt{scalarCLASS},\footnote{Our code will be made public upon completion of current work in progress.} which is based on \texttt{Axi-CLASS}~\cite{Poulin:2018dzj,Smith:2019ihp}, a modified version of cosmological Boltzmann solver \texttt{CLASS}~\cite{Lesgourgues:2011re,Blas:2011rf}.

For masses $m_\phi \gtrsim 10^{-26}~\mathrm{eV}$,
VFCs only impact the BBN era, so our CMB/BAO analyses simplify to constraining the deviation of the primordial helium abundance inferred from BBN.
For masses $m_\phi \lesssim 10^{-30}~\mathrm{eV}$, the ULDM field remains fixed on its potential and thus generates a constant variation that persists post recombination, recovering the scenario that has been previously studied~\cite{Hart:2019dxi,Sekiguchi:2020teg,toda_constraints_2025,Hart:2021kad,Seto:2022xgx,Lee:2022gzh,Schoneberg:2021qvd,Zhang:2022ujw,Schoneberg:2024ynd,Baryakhtar:2024rky,Baryakhtar:2025uxs,ACT:2025tim}.
For intermediate masses, VFCs alter the primordial helium abundance and can produce oscillatory changes to the process of recombination.
In any case, the amplitude of oscillations decays over time, resulting in negligible deviations of the fundamental constants in the late Universe that maintain compatibility with terrestrial constraints~\cite{Baryakhtar:2024rky,Baryakhtar:2025uxs}.
We note that there are additional constraints in our mass regime of interest that arise from small-scale structure observations, but we leave a combined study to future work.

We also account for changes in the evolution of the ULDM field due to the non-gravitational interactions with the SM.
The interactions induce a thermal contribution to the mass of the ULDM field, which non-trivially modifies its evolution and energy density.
This effect is relevant at the high end of the ULDM mass range we explore, where we are able to apply our simplified analysis of constraining VFCs through their modification of the primordial helium abundance; we infer the constraints on ULDM couplings to the SM from the resulting constraints on the helium abundance.

Finally, we explore the viability of our model of VFCs to address the Hubble tension.
Previous work suggests that oscillatory changes to the electron mass or fine structure constant during recombination can increase the inferred value of $H_0$ from CMB data~\cite{Lee:2022gzh}.
We do not find any significant increase in the value of $H_0$ for the range of ULDM masses that correspond to oscillatory VFCs during recombination; however, we note that our oscillation signature does not match well with the functional fit in Ref.~\cite{Lee:2022gzh}.
Comparing to previous work that considered constant VFCs~\cite{Seto:2022xgx,Schoneberg:2024ynd,Baryakhtar:2024rky,Baryakhtar:2025uxs}, which corresponds to our low ULDM mass regime, we find constraints of similar magnitude on the coupling constant.
We note that the shift to $H_0$ is slightly smaller compared to studies that considered the effects of constant VFCs during the CMB era only, because \emph{most} of those studies did not incorporate the impact of VFCs on the primordial helium abundance~\cite{Seto:2022xgx}.
The presence of ULDM perturbations in our study also impacts the inferred value of $H_0$.
While our result is specific to ULDM-induced VFCs, producing VFCs during recombination only may be quite challenging from a model-building perspective.

The structure of this paper is as follows.
We describe our model of ULDM-induced VFCs in Section~\ref{sec:model}.
We detail the impact of VFCs on the primordial helium abundance during BBN in Section~\ref{sec:bbn} and the impact of both the ULDM field and VFCs on the CMB anisotropies in Section~\ref{sec:cmb}.
In Section~\ref{sec:mcmc}, we describe our analysis methods and present our results.
We conclude in Section~\ref{sec:conclusion}.
We provide details of our modifications incorporated into \texttt{scalarCLASS} in Appendix~\ref{app:classbbn}. We analyze the effects of thermal mass on our results in Appendix~\ref{app:mcmctherm}
and showcase supplementary plots for our analyses in Appendix~\ref{app:traingle_plots}.

%%%%%%%%%%%%%%%%%%%%%%%%%%%%%%%%%%%%%%%%%%%%%%%%%%%%%%%%%%%%%%%%%%%%%%%%%%%%%%%
\section{Model}
\label{sec:model}
We represent ULDM as a scalar field $\phi$, which couples to the CP-even SM operators. 
These couplings can be linear, $\frac{\phi}{\Lambda}\mathcal{O}_{\mathrm{SM}}$, or quadratic, $\frac{\phi^2}{\Lambda^2}{\cal O}_{\rm SM}$, where $\mathcal{O}_\mathrm{SM}$ is a term in the SM Lagrangian and $\Lambda$ is some high scale at which new physics appears. 
Since linear couplings generate new, long-range forces that are strongly constrained by fifth-force experiments~\cite{EotWash_EP_2008,EotWash2020,HUST2021,IUPUI2016}, we focus on quadratic couplings in this work.
Such a model can be constructed by charging the scalar field under a $\mathbb{Z}_2$ symmetry, such that the leading coupling is quadratic and can arise from UV completions discussed in, e.g., Refs.~\cite{Brzeminski:2020uhm,Beadle:2023flm,Gan:2025nlu,Delaunay:2025pho,Banerjee:2025uwn,Becker:2025pgb}. 
Following the conventions of Refs.~\cite{Damour:2010rm,Damour:2010rp,Hees:2018fpg}, $\phi$ interacts with the Standard Model (SM) through the following Lagrangian:
\begin{equation}
  {\cal L}\supset 2\pi\frac{\phi^2}{M_{\rm pl}^2}\left[\frac{d_\alpha^{(2)}}{4e^2}F_{\mu\nu}F^{\mu\nu}-\frac{d_g^{(2)} \beta_3}{2 g_3}G^A_{\mu\nu}G^{A\mu\nu}-d_{m_e}^{(2)}m_e\bar e e-\sum_{i=u,d}\left(d_{m_i}^{(2)}+\gamma_{m_i}d_g^{(2)}\right)m_i\bar\psi_i\psi_i\right]\, ,
  \label{eq:Ldamour}
\end{equation}
where $M_{\rm pl}=1.22\times 10^{19}$ GeV is the Planck mass, $\beta_3$ is the QCD beta function, and $\gamma_{m_i}$ are the anomalous dimensions of the $u$ and $d$ quarks. 
The superscript $(2)$ makes explicit the quadratic nature of the scalar couplings. The mass term of the ULDM is given by $V(\phi)=\frac{1}{2}m_\phi^2\phi^2$. 

For this work, we focus on the couplings to the electron mass, $d_{m_e}^{(2)}$, and the photon field strength, $d_\alpha^{(2)}$.\footnote{The coupling to photon field strength is sometimes denoted by $d_e^{(2)}$ in the literature.}
We will explore the cosmological impact of the quark and gluon couplings in future work.
The interactions of interest in Eq.~\eqref{eq:Ldamour} induce a $\phi$-dependent variation of the electron mass and fine structure constant:
\begin{align}
\label{eq:mevar}
  \dfrac{\Delta m_e(a)}{m_e} &= 2\pi d_{m_e}^{(2)}\dfrac{\phi^2(a)}{M_{\rm pl}^2}\;, \\
\label{eq:alphavar}
  \dfrac{\Delta\alpha(a)}{\alpha} &= 2\pi d_\alpha^{(2)}\dfrac{\phi^2(a)}{M_{\rm pl}^2}\;,
\end{align}
where the dependence on the scale factor $a$ is written explicitly to emphasize the time-dependence of the variations.

Importantly, the interaction terms in Eq.~\eqref{eq:Ldamour} also generate a thermal contribution to the mass of the $\phi$ field. 
The induced thermal mass $m_\mathrm{th}$ is calculated using thermal field theory~\cite{Quiros:1999jp}, arising from loop corrections to the 2-point function associated with the $m_\phi^2 \phi^2$ term of the Lagrangian.
Thus, the square of the effective mass of $\phi$ is $m_{\rm eff}^2(T) = m_\phi^2 + m_{\rm th}^2(T)$, where $T$ is the temperature of the photon bath and $m_{\rm th}^2(T) = m_{{\rm th},m_e}^2(T) + m_{{\rm th},\alpha}^2(T)$ includes the contributions~\cite{Bouley:2022eer}
\begin{align}
  \label{eq:thmassme}
  m_{{\rm th},m_e}^2 (T) &= \dfrac{2\pi d_{m_e}^{(2)}}{M_{\rm Pl}^2}\dfrac{4m_e^2}{\pi^2}T^2\int_{m_e/T}^\infty dx \dfrac{\sqrt{x^2 -(m_e/T)^2}}{e^x + 1} \\
  \label{eq:thmassgamma}
  m_{{\rm th},\alpha}^2 (T) &\simeq \dfrac{2\pi d_{\alpha}^{(2)}}{M_{\rm Pl}^2} \dfrac{\alpha}{4\pi}\dfrac{\pi^2}{3}T^4
\end{align}
from the couplings with the electrons and photons, respectively.
The contribution $m_{{\rm th},m_e}^2$ is generated at one loop, while $m_{{\rm th},\alpha}^2$ is nonzero only at two loops.
This latter quantity is difficult to calculate due to overlapping UV divergences, so we use an estimate that is valid in the high-temperature limit ($T \gg m_e$); this choice sets an upper limit on the thermal mass contribution for $T \lesssim m_e$.
We provide details of the numerical implementation of the thermal masses in \texttt{scalarCLASS} in Appendix~\ref{app:classbbn}.

In the presence of interactions, the equation of motion of the ULDM scalar field is
\begin{equation}
\label{eq:eqm}
  \ddot{\phi} + 2aH\dot{\phi} + a^2m_{\rm eff}^2\phi = 0\;,
\end{equation}
where the overdot denotes a derivative with respect to conformal time $\eta$ and $H \equiv \dot{a}/a^2$ is the Hubble expansion rate.
The thermal mass term fully takes into account the energy transferred to the $\phi$ sector from the SM bath.
The thermal mass dictates the field evolution at early times; thus, the energy exchange compared to the intrinsic energy of the scalar field can be quite significant.
On the other hand, for the coupling to photons, the energy change in the photon bath is insignificant.
ULDM constitutes a small fraction of the total density (even in the presence of couplings with the SM) compared to the photon bath deep in the radiation domination.
Therefore, we can safely ignore the energy exchange between the $\phi$ field and the SM bath.
In the case of $m_e$ variation, the energy exchange is between the ULDM and the baryonic sector.
This effect is also small, since the contribution of electrons to the baryon energy density $\omega_b$ is negligible compared to nucleons.
Therefore, we can safely ignore the background-level energy transfer from ULDM to the SM sector in this work.

We assume that the ULDM abundance is generated via the misalignment mechanism.
The field $\phi$ is cosmologically frozen at its initial value when $H(a) \gg m_\mathrm{eff}$ and behaves as a cosmological constant with an equation of state $w = -1$.
When $H(a_\mathrm{osc}) = m_\mathrm{eff}$ at scale factor $a_\mathrm{osc}$, the field begins to oscillate.
Deep in the oscillation region where $H(a) \ll m_\mathrm{th}(a) < m_\phi$, $\phi$ acts as a dark matter component with an average equation of state $w = 0$, and its field value scales as $\phi \sim a^{-3/2}$. In the intermediate region $H(a) < m_\phi < m_\mathrm{th}(a)$, the thermal mass dominates the field evolution and the ULDM energy density redshifts faster than matter. The ULDM evolution in this range is non-trivial and is not captured by standard treatment of effective fluid approximations~\cite{Passaglia:2022bcr,Liu:2024yne,Cookmeyer:2019rna,Baryakhtar:2024rky,Urena-Lopez:2015gur,Cedeno:2017sou,Urena-Lopez:2023ngt,Moss:2025ymr}. The field in this region approximately scales as $\phi \sim m_{\rm eff}(a)^{-1/2}a^{-3/2}$~\cite{Sibiryakov:2020eir,Bouley:2022eer}, which redshifts faster than matter due to the scale factor (temperature) dependence of the $m_{\rm eff}$.

\begin{figure}[t]
  \centering
  \includegraphics[width=0.49\linewidth]{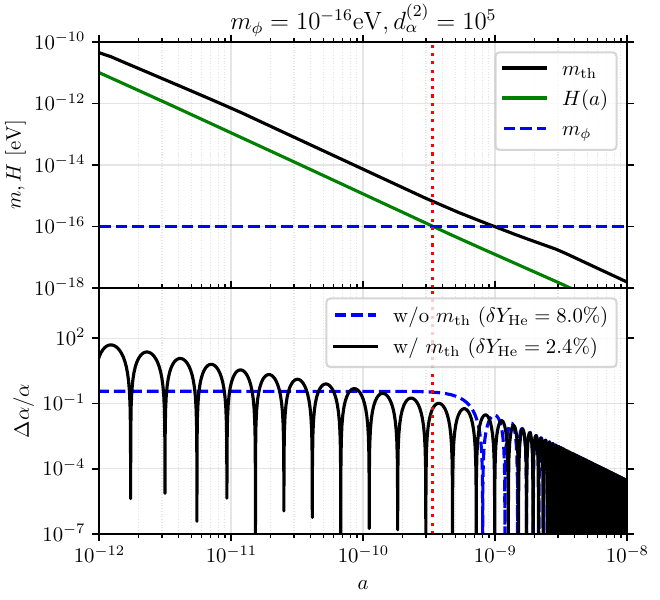}
  \includegraphics[width=0.49\linewidth]{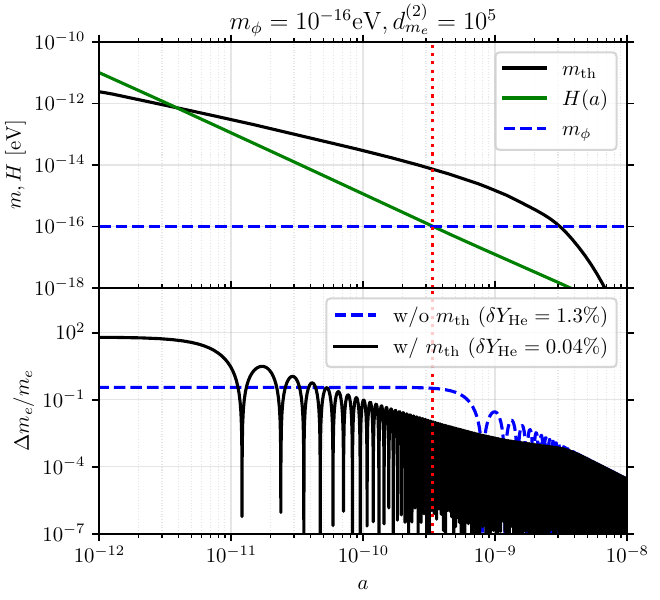}
  \caption{Evolution of $m_{\rm th}$ (top) and amount of variation (bottom) for $\alpha$ (left) and $m_e$ (right) with respect to the scale factor $a$.
    We set $m_\phi = 10^{-16}~\mathrm{eV}$ and $d_\alpha^{(2)} = 10^5$ (left) or $d_{m_e}^{(2)} = 10^5$ (right), and the associated variation of the primordial helium-4 abundance $\delta Y_\mathrm{He}$ is provided in the legends.
    For the $\alpha$ variation (left), the thermal mass dominates the evolution at early times when $m_{\rm th} > H, m_\phi$, and oscillations are always in effect.
    For the $m_e$ variation (right), oscillations begin once $m_\mathrm{eff} > H$.
    In either case, once $H \ll m_\mathrm{th} < m_\phi$, oscillations match onto the case in which the thermal mass is ignored (blue dashed), for which the onset of oscillations is much later, close to $H = m_\phi$ (red vertical dotted).
    }
  \label{fig:thmass}
\end{figure}

In Fig.~\ref{fig:thmass}, we demonstrate the effects of the thermal mass on the evolution of the VFCs, and we choose the initial field value such that $\phi$ constitutes the total dark matter abundance.
If we neglect the contribution from the thermal mass (\textcolor{blue}{blue} dashed lines), $\phi$ begins oscillating around when $H(a_\mathrm{osc}) = m_\phi$, and the value of $a_\mathrm{osc}$ is denoted by the vertical \textcolor{red}{red} dotted line.
Properly including the thermal mass (black solid lines) associated with the couplings $d_\alpha^{(2)}$ (left panel) and $d_{m_e}^{(2)}$ (right panel), we show that the effective mass $m_\mathrm{eff}$ is much larger than $m_\phi$ and thus the oscillations actually begin at much earlier times.

For the variation of $\alpha$ in Fig.~\ref{fig:thmass} (left panel), both $m_{{\rm th},\alpha}$ and $H$ scale as $T^2$ at early times.
The field $\phi$ is always oscillating (for the chosen set of example model parameters) due to $m_{{\rm th},\alpha} > H$ initially and $m_\phi > H$ later.
For the variation of $m_e$, the evolution of $m_{{\rm th},m_e}$ has a more complicated temperature dependence (right panel).
From Eq.~\eqref{eq:thmassme}, the thermal mass $m_{{\rm th},m_e}$ scales as $T$ for $T \gg m_e$ and falls exponentially in $T$ for $T \ll m_e$.
The resulting $\phi$ evolution is non-trivial during the period when the thermal mass is the dominant contribution to $m_\mathrm{eff}$.

\section{Impact on BBN}
\label{sec:bbn}
In this section, we consider the impact of time-dependent VFCs during the BBN era.
We review the standard BBN scenario, following the discussion in Ref.~\cite{Mukhanov:2003xs}, and summarize previous work on the effect of ULDM-induced VFCs~\cite{Sibiryakov:2020eir,Bouley:2022eer}, noting the different formulations we make to facilitate numerical calculations within \texttt{scalarCLASS}.

VFCs can modify the primordial abundances of light elements, such as helium, deuterium, and lithium.
Our main cosmological observable of interest is the anisotropy of the CMB, for which the only primordial element abundance of practical relevance is helium-4.
Thus, we focus on modifications of the helium-4 mass fraction $Y_\mathrm{He} \equiv \rho_{\mathrm{He}}/\rho_b$, the ratio of the energy density of helium-4 to the energy density of baryons.

For BBN calculations, it is convenient to use the concentration $X_A \equiv A n_A/n_N$ for a species with atomic number $A$ and number density $n_A$.
The total nucleon number density at early times is $n_N = n_n + n_p$, where $n_n$ and $n_p$ are the number densities of free neutrons and free protons, respectively.
The final helium-4 concentration $X_\mathrm{He}$ from BBN calculations and the mass fraction $Y_\mathrm{He}$ needed for CMB calculations differ at the sub-percent level, due to the binding energy of helium.
We neglect this small difference and approximate $X_\mathrm{He} \approx Y_\mathrm{He}$.

\subsection{Standard BBN}
We first review the standard BBN calculation of the helium-4 abundance, following Ref.~\cite{Mukhanov:2003xs}.
At early times, prior to the onset of BBN, the weak-interaction processes
\begin{equation}
  \label{eq:weak-process}
  n + \nu_e \leftrightarrow  p^+ + e^-, \quad n + e^+ \leftrightarrow p^+ + \bar{\nu}_e
\end{equation}
are efficient and keep the concentrations of neutrons and protons at their equilibrium values, given by
\begin{align}
  \label{eq:eqabandance}
  X_n^{\rm eq}(T) &= \dfrac{1}{1+e^{m_{np}/T}}\;,\\
  X_p^{\rm eq}(T) &= 1 - X_n^{\rm eq} = \dfrac{1}{1+e^{-m_{np}/T}}\;,
\end{align}
respectively, where $m_{np} \equiv m_n - m_p$ is the neutron-proton mass difference.
Weak interactions become inefficient as the SM bath temperature drops below $\sim \mathrm{MeV}$, causing the neutrons and protons to fall out of chemical equilibrium.
The relative abundance of neutrons and protons freezes out around the temperature $T_W \approx 0.75~{\rm MeV}$~\cite{Stadnik:2015kia,Rubakov:2017xzr},
and the process of weak freeze-out is described by the following Boltzmann equation:
\begin{equation}
  \label{eq:rateeqn}
  \dfrac{dX_n}{d\log a} = -\dfrac{\lambda_{np}}{H}\left(1+e^{-m_{np}/T}\right)(X_n - X_n^{\rm eq})\;.
\end{equation}
The total neutron-to-proton reaction rate from the two processes in Eq.~\eqref{eq:weak-process} is~\cite{Coc:2006sx}
\begin{equation}
  \label{eq:nprate}
  \lambda_{np} = \dfrac{1 + 3g_{A_n}^2}{\pi^3}G_F^2T^5J\left(m_{np} \over T\right)\;,
\end{equation}
where $G_F$ is the Fermi constant, $g_{A_n}$ is the neutron weak axial coupling, and 
\begin{equation}
  \label{eq:Jfac}
  J(x) = \dfrac{45\zeta(5)}{2} + \dfrac{21\zeta(4)}{2}x + \dfrac{3\zeta(3)}{2}\left(1-\dfrac{m_e^2}{2m_{np}^2}\right)x^2
\end{equation}
is a phase space factor and $\zeta$ is the Riemann zeta function.

After the weak freeze-out, the neutron concentration continues to decrease due to neutron decay:
\begin{equation}
  \label{eq:neutron-decay-abundance}
  \dfrac{dX_n}{d\log a} = -\dfrac{X_n \Gamma_n}{H}\;,
\end{equation}
where $\Gamma_n$ is the inverse neutron lifetime
\begin{equation}
  \label{eq:nudecay}
  \Gamma_n = \dfrac{1 + 3g_{A_n}^2}{2\pi^3}G_F^2T^5P\left(m_{np} \over m_e\right)\;,
\end{equation}
with a phase space factor given by
\begin{equation}
  \label{eq:phase-space}
  P(x) = {1 \over 60}\left[(2x^4 - 9x^2-8)\sqrt{x^2 -1} + 15x\ln(x+\sqrt{x^2-1})\right]\;.
\end{equation}
Solving Eq.~\eqref{eq:neutron-decay-abundance}, the resulting neutron concentration at a time after weak freeze-out is
\begin{equation}
  \label{eq:nuabbbn}
  X_{n} (a_D) = X_{n}(a_W) \exp\left(-\int_{a_W}^{a_D} \dfrac{\Gamma_n}{H}d\log a \right)\;,
\end{equation}
where $X_{n}(a_W)$ is the value of neutron concentration at weak freeze-out $a = a_W$, and we integrate up to a scale factor $a_D > a_W$.
The formation of helium-4 relies on the conversion of deuterium nuclei into heavier elements, which becomes efficient when the deuterium bottleneck opens wide around the temperature $T_D \approx B_D/30 \approx 0.075~\mathrm{MeV}$, where $B_D$ is the deuterium binding energy, and the corresponding scale factor $a_D$.
Equation~\eqref{eq:nuabbbn} holds up to this point, when most of the free neutrons become bound within helium-4 nuclei; thus, the resulting primordial helium-4 concentration and mass fraction relevant for CMB calculations are given by
\begin{equation}
  \label{eq:Yp_sm}
  Y_\mathrm{He} \approx X_\mathrm{He} \approx 2X_n(a_D)\;.
\end{equation}
This expression produces $Y_\mathrm{He} \approx 0.25$, close to the SM prediction $Y_\mathrm{He}^{\rm SM} = 0.245 \pm 0.003$~\cite{ParticleDataGroup:2024cfk}, which properly takes into account all processes relevant for BBN.

\subsection{Modified BBN from VFCs}
The standard BBN reaction rates and evolution equations in the previous subsection are modified in the presence of ULDM-induced VFCs~\cite{Sibiryakov:2020eir,Bouley:2022eer}, which in turn modifies the BBN prediction for $Y_\mathrm{He}$. We can parameterize the neutron-proton mass difference as
\begin{equation}
  \label{eq:np_mass_diff}
  m_{np} = m_n - m_p \approx b\alpha\Lambda_{\rm QCD} + (m_d - m_u)\;,
\end{equation}
where $b$ is a constant such that $b\alpha\Lambda_{\rm QCD} \approx -0.76~{\rm MeV}$~\cite{Coc:2006sx},
which results in the variation
\begin{equation}
  \label{eq:mnpvar}
  \dfrac{\Delta m_{np}}{m_{np}} = \left(\dfrac{b\alpha\Lambda_{\rm QCD}}{m_{np}}\right) 2\pi d_\alpha^{(2)} \dfrac{\phi^2}{M_{\rm Pl}^2} \;.
\end{equation}
Modifying $m_{np}$ affects the evolution of the neutron concentration and thus the value of $Y_\mathrm{He}$.
To mitigate any systematic errors originating from the small disagreement between our simplified calculation in Eq.~\eqref{eq:Yp_sm} and the full derivation, we compute the fractional change $\delta Y_\mathrm{He} \equiv \Delta Y_\mathrm{He}/Y_\mathrm{He}$ in our formalism.
We then apply those corrections to $Y_{\rm He}^{\rm SM}$ to get the modified $Y_{\rm He}$ in the presence of VFCs.
For the purposes of numerical evaluation of $\Delta X_n$, it is convenient to find the variation for the differential equation in Eq.~\eqref{eq:rateeqn}, rather than using the equivalent integrated expression for the neutron concentration in Ref.~\cite{Bouley:2022eer}.
We obtain
\begin{equation}
  \label{eq:Drateeq}
  \dfrac{d\Delta X_n}{d\log a} = -{X_n \over H}\left[\dfrac{\Delta\lambda_{np}}{\lambda_{np}}  -\dfrac{e^{-m_{np}/T}}{1+e^{-m_{np}/T}}\dfrac{\Delta m_{np}}{T}+ \dfrac{\Delta X_n - \Delta X_n^{\rm eq}}{X_n - X_n^{\rm eq}}\right]\;,
\end{equation}
where~\cite{Bouley:2022eer}
\begin{align}
\label{eq:DXneq}
  \Delta X_{\rm eq} &= -\dfrac{m_{np}}{2T[1+ \cosh(m_{np}/T)]}\dfrac{\Delta m_{np}}{m_{np}}\;,\\
  \label{eq:Dlnp}\dfrac{\Delta\lambda_{np}}{\lambda_{np}} &=  \dfrac{\Delta J\left(m_{np} / T\right)}{J\left(m_{np} / T\right)} = \dfrac{m_{np}}{T}\dfrac{J'}{J}\dfrac{\Delta m_{np}}{m_{np}} - \dfrac{3\zeta(3)}{2J}\dfrac{m_e^2}{T^2}\left(\dfrac{\Delta m_e}{m_e} - \dfrac{\Delta m_{np}}{m_{np}}\right)\, ,
\end{align}
are the variation of the equilibrium neutron concentration and the relative variation of the neutron-to-proton reaction rate, respectively, and $J'(x) = dJ/dx$.
We numerically solve Eq.~\eqref{eq:Drateeq} until weak freeze-out to obtain the variation of the neutron concentration at weak freeze-out, $\Delta X_n(a_W)$.
From the variation of Eq.~\eqref{eq:nuabbbn}, the corresponding change in the neutron concentration is
\begin{equation}
  \label{eq:varXnbbn}
  \dfrac{\Delta X_{n}}{X_{n}} (a_D) = \dfrac{\Delta X_{n}}{X_{n}} (a_W) -\int_{a_W}^{a_D}\dfrac{\Delta\Gamma_n}{H}d\log a\ - \dfrac{\Gamma_n}{H}\left.\dfrac{\Delta a}{a}\right\vert_{a_W}^{a_D}\;.
\end{equation}
The last term on the right-hand side of Eq.~\eqref{eq:varXnbbn} is related to the variation in temperature:
\begin{equation}
  \label{eq:varTbbn}
  \dfrac{\Gamma_n}{H}\left.\dfrac{\Delta a}{a}\right\vert_{a_W}^{a_D} \approx -\left.\dfrac{\Gamma_n}{H} \dfrac{\Delta T}{T} \right\vert_{a_D}^{a_W} \approx 0 \;,
\end{equation}
for the VFCs we consider in this work.
The variation of $T_{\rm D}$ is related to the variation of the binding energy of deuterium $(\Delta B_D)$, which is zero in our case.
Since $\left.({\Gamma_n}/{H})\right\vert_{a_{W}} \ll 1$, the contribution at $a_W$ can be ignored.
Evaluating $\Delta\Gamma_n$ and using Eq.~\eqref{eq:Yp_sm}, the relative variation of $Y_\mathrm{He}$ is~\cite{Bouley:2022eer}
\begin{equation}
  \label{eq:DY_p}
  \delta Y_{\rm He} \equiv \dfrac{\Delta Y_\mathrm{He}}{Y_\mathrm{He}} = \dfrac{\Delta X_{n}}{X_{n}} (a_W) -  \int_{a_W}^{a_D} \dfrac{\Gamma_n}{H}\left[5{\Delta m_e \over m_e} - \dfrac{m_{np}}{m_e}\dfrac{P'}{P}\left(\dfrac{\Delta m_e}{m_e} - \dfrac{\Delta m_{np}}{m_{np}}\right)\right]d\log a \;,
\end{equation}
where $P'(x) = dP/dx$.
In Appendix~\ref{app:classbbn}, we provide the details of our numerical implementation of this expression into \texttt{scalarCLASS}.

In the legends of the Fig.~\ref{fig:thmass}, we provide the values of $\delta Y_\mathrm{He}$ following Eq.~\eqref{eq:DY_p}.
Through the variations of $\alpha$ and $m_e$, $\delta Y_\mathrm{He}$ depends on the evolution of $\phi$ between weak decoupling and BBN $(a\approx 10^{-10}-10^{-9})$.
Note that neglecting the effects of the thermal mass, discussed in Sec.~\ref{sec:model}, causes the field values (and thus the amount of variation in $\alpha$ or $m_e$) during this period to be larger than they should be. This, in turn, results in an artificially large effect on $\delta Y_\mathrm{He}$.

In this work, we consider only positive couplings, as negative couplings can lead to a tachyonic instability when the induced mass dominates. Although this can serve as a mechanism for DM production, the initial $\phi$ value needs to be fine-tuned to avoid overproduction.

\section{Impact on CMB anisotropy}
\label{sec:cmb}
In this section, we discuss the effects of ULDM-induced VFCs on the anisotropy of the CMB.
We first focus on how variations of $\alpha$ and $m_e$ impact the CMB.
Then, we briefly summarize the gravitational impact of pure ULDM (i.e., no SM couplings and thus no VFCs) on the CMB and discuss how previous ULDM analyses guide us in separating our study into two mass regimes.
Finally, we present the combined effects on the CMB power spectra from ULDM and its corresponding VFCs.

Throughout the section, we demonstrate changes to the CMB temperature (TT) and polarization (EE) power spectra, as well as other quantities relevant for CMB calculations.
Our comparisons are with respect to a fiducial $\Lambda$CDM cosmology, fixing the 6 standard $\Lambda$CDM parameters (reduced Hubble constant $h$, baryon energy density $\omega_b = \Omega_b h^2$, CDM energy density $\omega_{\rm CDM} = \Omega_{\rm CDM} h^2$, amplitude $A_s$ and spectral tilt $n_s$ of the primordial power spectrum, and optical depth to reionization $\tau_{\rm reio}$) to their best-fit \textit{Planck} 2018 TTTEEE+lensing values~\cite{Planck:2018vyg}.

\begin{figure}
    \centering
    \includegraphics[width=0.7\linewidth]{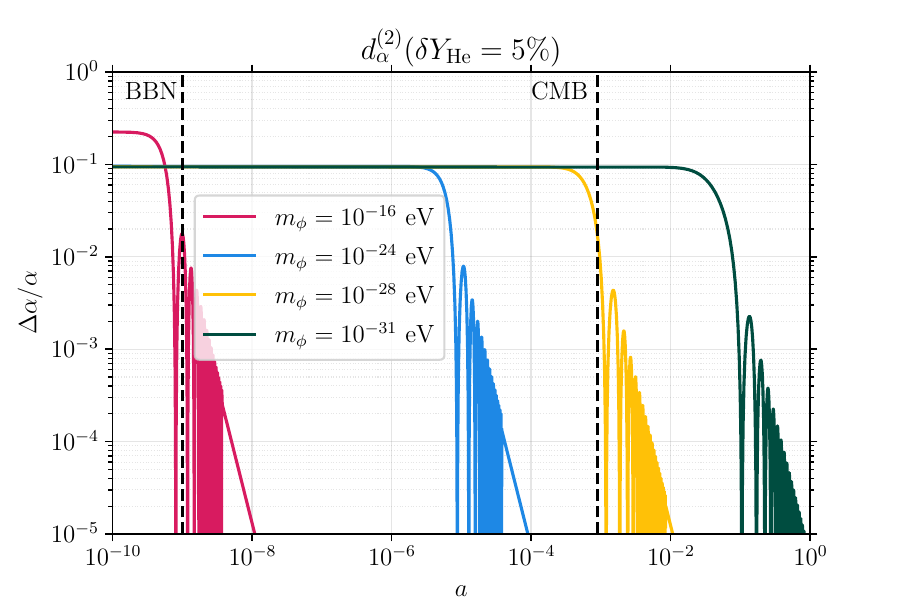}
    \caption{Evolution of the variation of $\alpha$ as a function of scale factor for different masses of $\phi$, neglecting the contribution of the thermal mass.
      We choose the amount of variation such that $\delta Y_\mathrm{He} = 5\%$ from the BBN era, and thus the mass $m_\phi$ determines how significant the variation is during the recombination era.
      Only the smaller values of $m_\phi$ affect recombination appreciably, with the case of $m_\phi = 10^{-28}~{\rm eV}$ exhibiting oscillations around recombination.
      Oscillations occur during BBN for $m_\phi = 10^{-16}~{\rm eV}$, requiring the variation at very small $a$ to be larger than the other cases in order to achieve $\delta Y_\mathrm{He} = 5\%$.}
    \label{fig:variation_w_diff_mass}
\end{figure}

\subsection{Effect of VFCs}

VFCs alter the CMB power spectra by modifying the time of recombination, the evolution of the ionization fraction, and the visibility function.
The process of recombination---and thus the resulting ionization fraction---is governed by atomic energy levels, transition rates, recombination and ionization rates, and radiative transfer physics, all of which exhibit certain scalings with $\alpha$ and $m_e$~\cite{Ali-Haimoud:2010hou,Ali-Haimoud:2010tlj,Chluba:2010ca,Chluba:2015gta,Lee:2020obi}.
For example, the time of recombination is sensitive to the energy levels of the hydrogen atom, which scale as $\propto \alpha^2 m_e$.
Additionally, the Thomson cross section $\sigma_\mathrm{T} \propto \alpha^2 / m_e^2$ controls the thermal evolution of baryons and enters into the calculation of the visibility function $g(\eta) = e^{-\tau}d\tau/d\eta$, where $\tau$ is the optical depth.
VFCs during recombination change these rates and modify the free electron fraction $x_e$.
They also shift the peak and width of the visibility function, which define the time $\eta_\ast$ and thickness, respectively, of the surface of last scattering.
A shift in $\eta_\ast$ changes the associated sound horizon and angular diameter distance, which in turn shifts the locations of the CMB acoustic peaks; it also changes the amplitude of the first acoustic peak via the early integrated Sachs–Wolfe effect.
Modifying the thickness of the last scattering surface alters the amount of diffusion damping, which affects the amplitude of the high-multipole acoustic peaks of the CMB.

VFCs also affect CMB anisotropies through modifications of the primordial helium-4 abundance, $Y_\mathrm{He}$~\cite{Trotta:2003xg,Schoneberg:2024ynd,Baryakhtar:2024rky}. Before helium-4 recombination, which occurs prior to H recombination, more ionized helium-4 (for a fixed $\omega_b$) corresponds to more free electrons, which increase diffusion damping and suppress the high-multipole CMB anisotropies. Thus, a larger $Y_\mathrm{He}$ suppresses the CMB TT power spectrum due to enhanced diffusion damping. 
A change in $Y_\mathrm{He}$ only mildly alters the redshift of decoupling $(z_{\rm dec})$ through its dependence on the available free electrons for H recombination. The CMB polarization spectrum has additional sensitivity to $Y_\mathrm{He}$ through the reionization process. The number density of reionized electrons scales as $\omega_b(1 - Y_\mathrm{He})$. Changing $Y_\mathrm{He}$ affects the redshift of reionization when the optical depth of reionization $\tau_{\rm reio}$ is kept fixed. The redshift dependence of the VFC during reionization may also strongly influence the dynamics of the reionization process~\cite{Trotta:2003xg}.

The VFCs we consider in this work depend on the value of the ULDM $\phi$ field, which evolves with time, as demonstrated in Fig.~\ref{fig:thmass}.
Neglecting the contribution from the thermal mass, the importance of VFCs during different cosmic epochs is controlled by the ULDM mass $m_\phi$, as we show in Fig.~\ref{fig:variation_w_diff_mass} for a variation of $\alpha$, fixing $d_\alpha^{(2)}$ such that $\delta Y_\mathrm{He} = 5\%$.
Both the onset and the period of the oscillations are set by $m_\phi$: larger values of $m_\phi$ correspond to a smaller period of oscillations that begin at earlier times, and the amplitude of oscillations decreases monotonically over time.

\begin{figure}[t]
    \centering
    \includegraphics[width=0.49\linewidth]{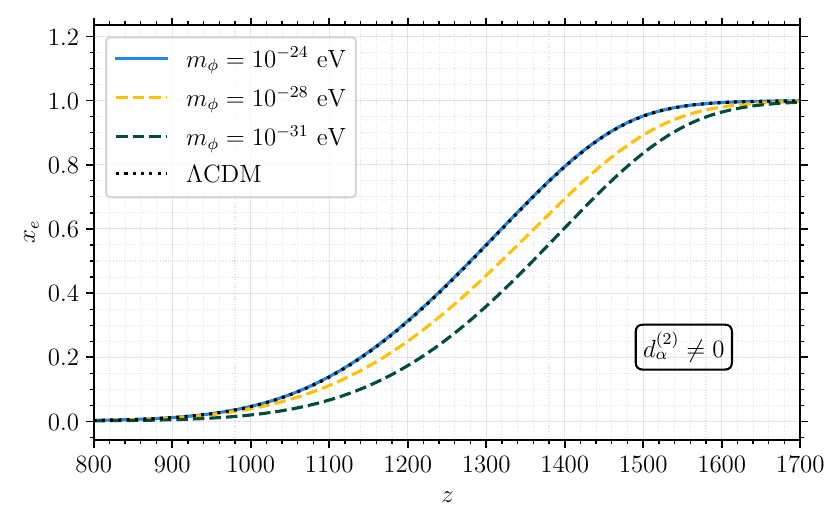}
    \includegraphics[width=0.49\linewidth]{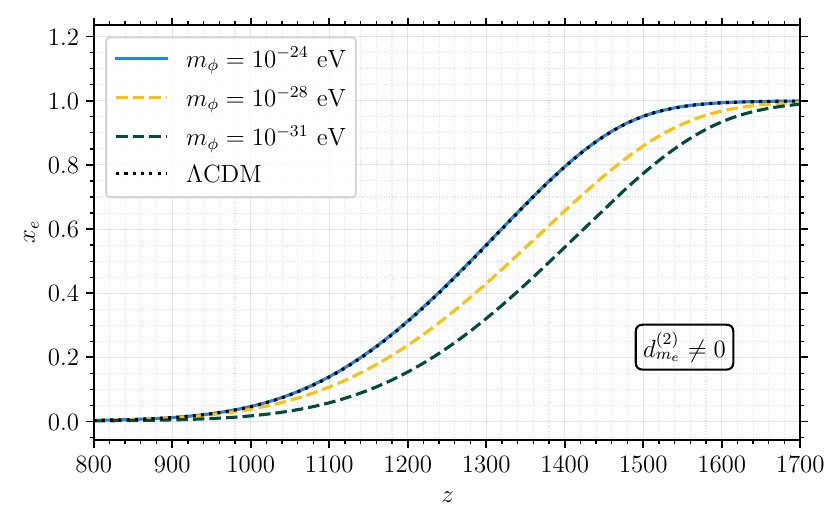}
    \includegraphics[width=0.49\linewidth]{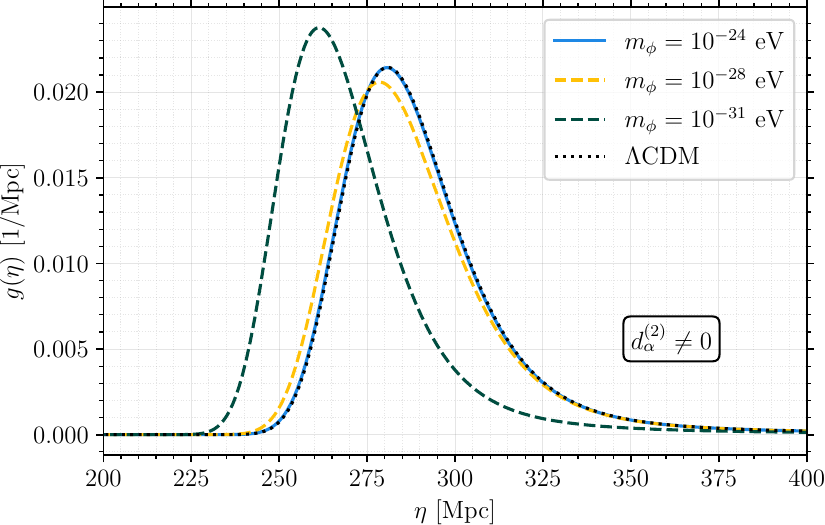}
    \includegraphics[width=0.49\linewidth]{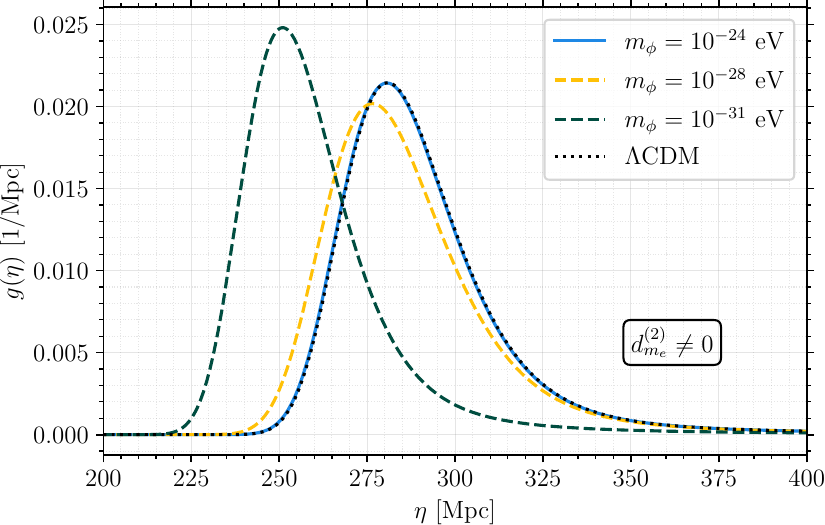}    
    \caption{Evolution of the free-electron fraction $x_e$ (top) and visibility function $g$ (bottom) in the presence of $\alpha$ (left) and $m_e$ (right) variations for three different masses of $\phi$, aligning with those presented in Fig.~\ref{fig:variation_w_diff_mass}.
      We neglect the contribution of the thermal mass.
      The specific values of the ULDM coupling and fraction match those in Fig.~\ref{fig:Cl_low_mass}, enforcing $\delta Y_\mathrm{He} = 5\%$ such that all $x_e$ curves align at high redshift.
      We plot the visibility function with respect to conformal time, because its width corresponds to the comoving damping scale.
      We also show the corresponding curves for $\Lambda$CDM (black dotted line).
      For $m_\phi = 10^{-24}$ eV, the VFC has a negligible impact on recombination.
      For smaller masses, the VFC alters both the peak location and width of the visibility function.
      }
    \label{fig:xegz}
\end{figure}

For the highest mass of $m_\phi = 10^{-16}$ eV (\textcolor{myred}{red}) in Fig.~\ref{fig:variation_w_diff_mass}, the oscillations begin during BBN, modifying the value of $Y_\mathrm{He}$; the amplitude of oscillations decays as $\phi^2 \sim a^{-3}$, resulting in a negligible variation by the time of recombination.
For $m_\phi = 10^{-24}$ eV (\textcolor{myblue}{blue}), $\phi$ is frozen in its potential during BBN, resulting in a modification to $Y_\mathrm{He}$ that is induced by an effective constant shift in $\alpha$ during BBN, as in Ref.~\cite{Dent:2007zu}.
Again, the VFC during recombination is negligible, and the only impact on the CMB is via the change in $Y_\mathrm{He}$.

For the lower masses in Fig.~\ref{fig:variation_w_diff_mass}, the presence of VFCs can more directly affect the recombination era.
For $m_\phi = 10^{-28}$~eV (\textcolor{myyellow}{yellow}), the field oscillations begin slightly before the onset of recombination, resulting in a variation that is slightly smaller at recombination than at BBN.
For $m_\phi = 10^{-31}$ eV (\textcolor{mygreen}{green}), there is a constant variation during recombination, after which the field begins to oscillate.
In both cases, the field is frozen during BBN, producing a constant variation that modifies $Y_\mathrm{He}$.
Therefore, while a constant variation during recombination is similar to previous CMB studies with a constant VFC~\cite{Kaplinghat:1998ry,Hannestad:1998xp,Battye:2000ds,Avelino:2000ea,Avelino:2001nr,Martins:2003pe,scóccola2009wmap5yearconstraintstime,Martins:2010gu,Menegoni:2012tq,Planck:2014ylh,Hart:2017ndk,Hart:2019dxi,Sekiguchi:2020teg,Hart:2022agu,Tohfa:2023zip,Baryakhtar:2024rky,Baryakhtar:2025uxs,Schoneberg:2024ynd}, only a few studies have accounted for an accompanying modification to $Y_\mathrm{He}$~\cite{Seto:2022xgx,Schoneberg:2024ynd,Baryakhtar:2024rky,Baryakhtar:2025uxs}.

In Fig.~\ref{fig:xegz}, we show the evolution of the free-electron fraction $x_e$ and the visibility function $g(\eta)$ for the same values of the ULDM mass $m_\phi$ used in Fig.~\ref{fig:variation_w_diff_mass}.
For $m_\phi = 10^{-24}$ eV, since the VFC is negligible at recombination, we recover the same $x_e$ and $g(\eta)$ evolution as in $\Lambda$CDM.
For $m_\phi = 10^{-28}$ eV, VFCs slightly shift the peak of the visibility function to earlier times, which results in an earlier recombination.
It also slightly broadens the width of the visibility function, which slightly increases the amount of diffusion damping~\cite{Kaplinghat:1998ry}.
Moreover, since the oscillatory VFCs progressively diminish as recombination progresses, they alter the visibility function and $x_e$ more at higher redshifts, at the onset of recombination.
For $m_\phi = 10^{-31}$ eV, on the other hand, VFCs cause a substantial shift of the visibility function to earlier times.
In this case, the width of the visibility function is narrower, reducing the amount of diffusing damping.
We defer the detailed description of how the altered evolution of $x_e$ and $g(\eta)$ modify the CMB power spectra to Sec.~\ref{sec:cmb-combined}.

\begin{figure}
    \centering
    \includegraphics[width=0.49\linewidth]{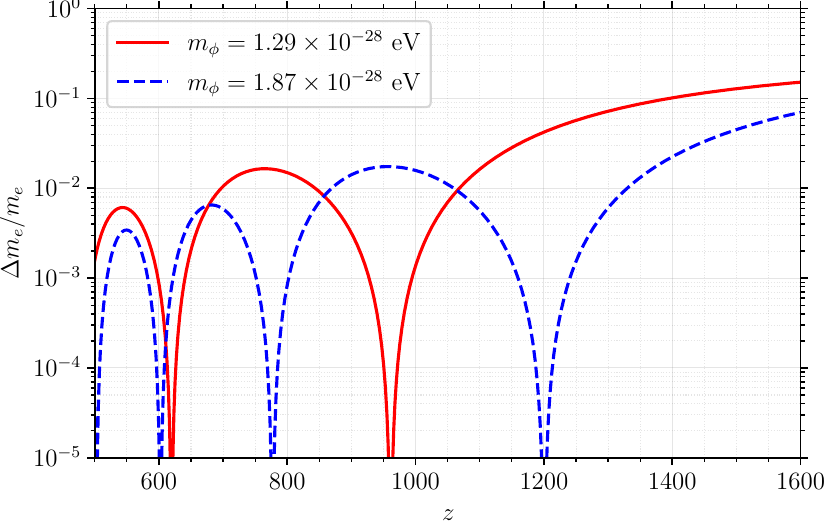}
    \includegraphics[width = 0.49\linewidth]{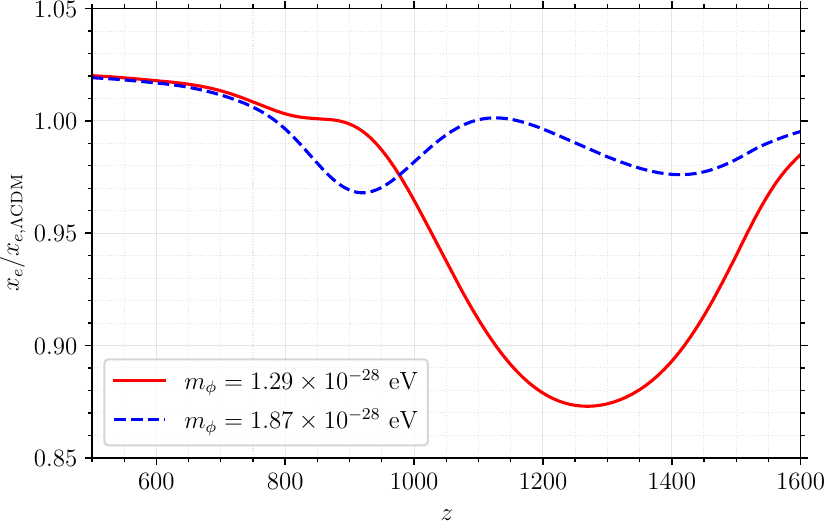}
    \caption{Oscillatory variation of $m_e$ (left) and its effect on the free-electron fraction $x_e$, with respect to $\Lambda$CDM (right), for two values of the mass of $\phi$ that correspond to oscillations starting around the onset of recombination.
      We neglect the contribution of the thermal mass and fix the ULDM coupling and fraction to achieve $\delta Y_{\rm He} = 5\%$.
      We note that similar oscillatory behaviors also appear for the $\phi$-induced variation of $\alpha$.
    }
    \label{fig:Dxe}
\end{figure}

For a range of masses near $m_\phi \sim 10^{-28}$~eV, oscillatory VFCs during the epoch of recombination imprint oscillation features in $x_e$ and $g(\eta)$.
In Fig.~\ref{fig:Dxe}, we show the oscillatory variation in $m_e$ and the resulting change to $x_e$ for two masses slightly above $10^{-28}$~eV.
The case with the larger mass generates more oscillations, but since the oscillations begin earlier, the amplitude is weaker at the time of recombination.
We note that an oscillatory-like $m_e$ (or $\alpha$) variation during recombination has been shown to produce a larger inferred value of the Hubble constant $H_0$ from a CMB analysis, drawing connections to the Hubble tension~\cite{Lee:2022gzh}.
However, the data-driven features of VFCs in Ref.~\cite{Lee:2022gzh} are distinct from our model-driven oscillations; for example, our oscillation amplitude decreases with time, while those in Ref.~\cite{Lee:2022gzh} grow with time.
We explore the implications for $H_0$ in Sec.~\ref{sec:mcmc-hubble}.

\subsection{Gravitational Effect of ULDM}
In the previous section, we discussed the effects of VFC on the CMB anisotropies. However, the mere presence of the ULDM field itself can alter the CMB anisotropies away from the standard prediction from cold dark matter (CDM).
Before the onset of oscillations, $\phi$ behaves as a DE-like fluid, resulting in a background and perturbation evolution that differs from CDM.
These gravitational effects on the CMB have been studied extensively~\cite{Hu:2000ke,Amendola:2005ad,Marsh:2010wq,Marsh:2015xka}, and we briefly summarize the relevant phenomenology here, neglecting any non-gravitational coupling with the SM.

A coherently oscillating field $\phi$ manifests a new ``Jeans'' scale $k_J \sim \sqrt{m_\phi H}$ below which ULDM perturbations cannot cluster.
Thus, on scales smaller than $k_J$ (i.e., $k > k_J$), ULDM suppresses the growth of structure and the matter power spectrum.
The scale $k_J$ roughly corresponds to the mode entering the horizon during the onset of the $\phi$ oscillations, and pure ULDM behaves generically as CDM at larger scales.
If the scales probed by CMB experiments are larger than $k_J$ (i.e., $k < k_J$), we can treat ULDM as CDM in the CMB analyses.
For sufficiently small ULDM masses, $k_J$ is within the scales probed by current CMB experiments, and the CMB power spectra are modified due to the suppression of small-scale clustering and the period of DE-like background evolution.
As a result, ULDM is constrained to constitute only a part of the total dark matter energy density~\cite{Hlozek:2014lca,Hlozek:2017zzf,Lague:2021frh}.
We define the present-day fraction of ULDM as $f_\phi \equiv \Omega_\phi/\Omega_{\rm DM}$ and assume the remaining dark matter relic density is in CDM: $\Omega_\mathrm{CDM} = (1-f_\phi) \Omega_\mathrm{DM}$.

\subsection{Combined effect}
\label{sec:cmb-combined}

We now consider how the CMB power spectra change in the presence of ULDM with quadratic couplings to the SM, combining the impact of both the VFC and the gravitational effects induced by ULDM.
Building on our earlier discussion of ULDM without SM couplings, we can divide the analysis into two mass regimes.
If the scales probed by current CMB observations satisfy $k < k_J$, corresponding to ``high'' $\phi$ masses, ULDM behaves effectively as CDM.
In contrast, for ``low'' masses, there are observable CMB modes that satisfy $k > k_J$, in which case we must consider the full dynamical evolution of the ULDM field.
In the remainder of this section, we discuss how the addition of SM couplings impacts the CMB power spectra in both the low- and high-mass regions.
\subsubsection{High-mass region}

\begin{figure}[t]
    \centering
    \includegraphics[width=0.49\linewidth]{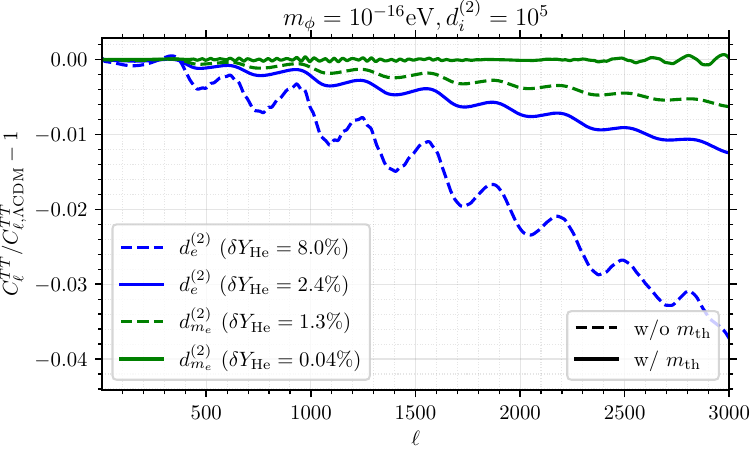}
    \includegraphics[width=0.49\linewidth]{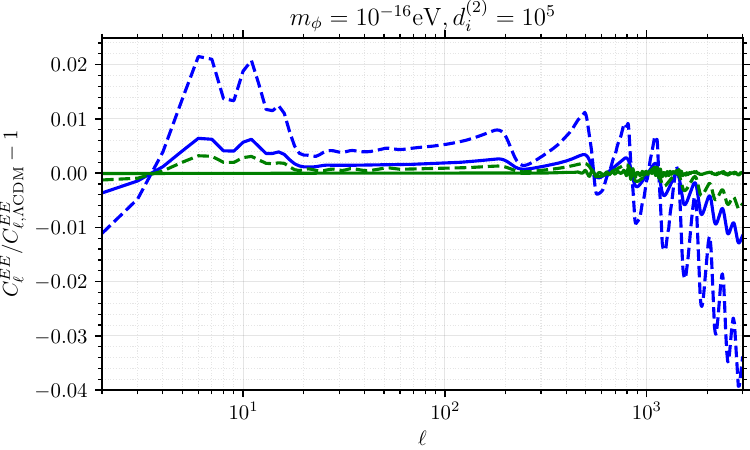}
    \caption{Residuals of the CMB TT (left) and EE (right) power spectra, with respect to $\Lambda$CDM, due to the modified value $Y_\mathrm{He}$ from VFCs.
      Solid (dashed) lines show the effects with (without) thermal mass.
      An increase in $\delta Y_\mathrm{He}$ suppresses the CMB tail due to increased diffusion damping.
      The large-scale features in EE result from the difference in the reionization visibility function.
      Properly accounting for $m_\mathrm{th}$ reduces the value of $Y_{\rm He}$ and lessens the impact on the CMB power spectra for the chosen set of parameters, indicated at the top of each panel.
    }
    \label{fig:Cl_high_mass}
\end{figure}

For $m_\phi \gtrsim 10^{-23}~{\rm eV}$, ULDM behaves like CDM for observed CMB scales and can thus constitute the total dark matter energy budget.
As demonstrated in Fig.~\ref{fig:variation_w_diff_mass}, the VFC for this mass range predominantly affects BBN, so the primary effect on the CMB is through the modification of $Y_\mathrm{He}$.
In Fig.~\ref{fig:Cl_high_mass}, we show the impact on the TT and EE power spectra due to changes in $Y_\mathrm{He}$ associated with ULDM-induced VFCs (as indicated in the legend) during BBN for $m_\phi = 10^{-16}~{\rm eV}$.
Larger values of $Y_\mathrm{He}$ result in increased diffusion damping at high multipoles $\ell$.
The large-scale changes in the EE power spectrum originate from the difference in the reionization bump: for a fixed $\omega_b$ and $\tau_{\rm reio}$, modifying $Y_\mathrm{He}$ changes the peak of the visibility function at reionization~\cite{Trotta:2003xg}.

Properly accounting for the effects of the thermal mass---for the chosen set of fixed parameters in Fig.~\ref{fig:Cl_high_mass}---leads to a smaller increase in the value of $Y_\mathrm{He}$ prediction, as demonstrated in Fig.~\ref{fig:thmass}.
Hence, the corresponding changes in the CMB spectrum are smaller.
In this case, we expect that including the thermal mass would result in a relaxed constraint on the VFC coupling.
However, this conclusion is not generally true: for slightly smaller $m_\phi$, the thermal mass can result in a larger increase in the value $Y_\mathrm{He}$.

\subsubsection{Low-mass region}

\begin{figure}
    \centering
    \includegraphics[width=0.49\linewidth]{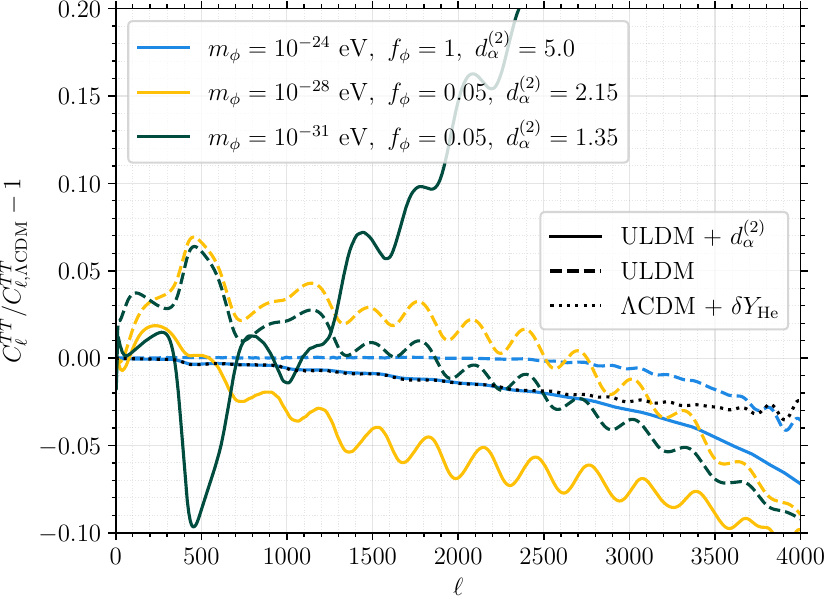}
    \includegraphics[width=0.49\linewidth]{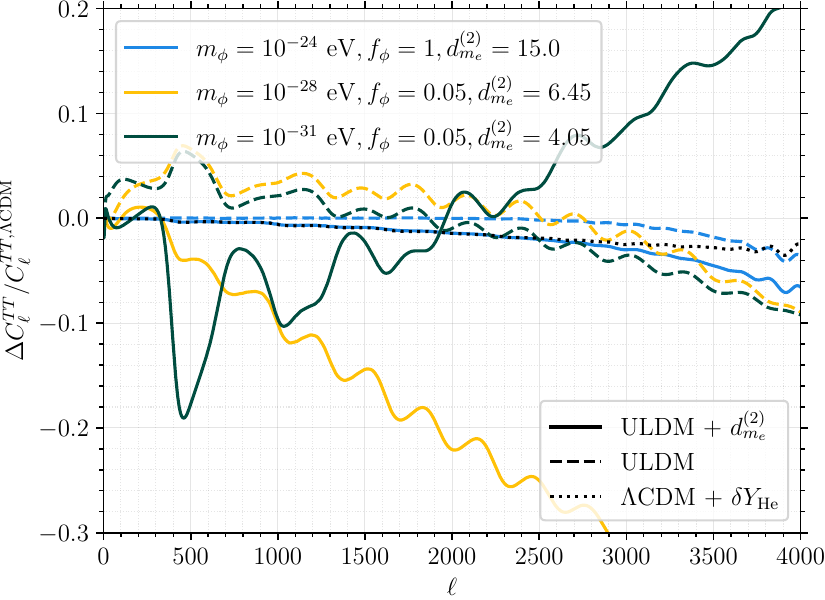}
    \caption{Residuals of the CMB TT power spectrum for a ULDM-induced variation of $\alpha$ (left) and $m_e$ (right) at different masses $m_\phi$, with respect to $\Lambda$CDM.
      The ULDM fraction of the total present-day dark matter density is denoted by $f_\phi$.
      The gravitational effects of ULDM alone (without the VFC) are shown in dashed lines, while the combined effects of ULDM and its associated VFC are shown in solid lines.
      For each line, the coupling (as indicated in the caption) is chosen such that $\delta Y_\mathrm{He} = 5\%$, so the impact due to the modified $Y_\mathrm{He}$ is the same for all masses. We neglect the contribution of the thermal mass in this mass range.
      The effects of VFCs on the power spectrum become progressively stronger for smaller masses, which have a more appreciable impact on recombination, as shown in Figs.~\ref{fig:variation_w_diff_mass} and \ref{fig:xegz}.
      For comparison, we show the residuals for $\Lambda$CDM with an additional $\delta Y_\mathrm{He} = 5\%$ as the black dotted line.
      }
    \label{fig:Cl_low_mass}
\end{figure}

At smaller masses ($m_\phi \lesssim 10^{-23} {\rm eV}$), CMB is sensitive to the evolution of ULDM.
In Fig.~\ref{fig:Cl_low_mass}, we show the impact on the TT power spectrum due to ULDM-induced VFCs.
For each plotted curve, we fix the value of $f_\phi$ to its approximate upper limit from a previous CMB analysis of ULDM with no VFCs~\cite{Hlozek:2014lca}, for demonstration purposes.
We also fix the VFC coupling such that $\delta Y_\mathrm{He} = 5\%$, rendering the impact due to the modified $Y_\mathrm{He}$ to be the same in all cases; thus, differences between the curves stem only from the ULDM evolution and the VFC during recombination.
We show the impact of the pure ULDM case in dashed lines and the case of ULDM with VFCs in solid lines.
The corresponding redshift-dependence of the VFC for each case is depicted in Fig.~\ref{fig:variation_w_diff_mass}.

For $m_\phi = 10^{-24}$~eV in Fig.~\ref{fig:Cl_low_mass}, pure ULDM suppresses the power spectrum at high multipoles, corresponding to the large associated value of $k_J$ (\textcolor{myblue}{dashed blue}).
We still expect $f_\phi \approx 1$ to be allowed by CMB data~\cite{Hlozek:2014lca,Hlozek:2017zzf,Lague:2021frh}, since the ULDM effects are mostly confined to the high-$\ell$ tail of the CMB.
The additional $\delta Y_\mathrm{He}$ contribution from the VFC produces further suppression of the CMB damping tail (\textcolor{myblue}{solid blue}).
Note that for this mass, the effect of the VFC at recombination is negligible, as shown in Fig.~\ref{fig:xegz}.

For $m_\phi = 10^{-28}$~eV in Fig.~\ref{fig:Cl_low_mass}, we fix $f_\phi = 0.05$, corresponding to the approximate upper limit in Ref.~\cite{Hlozek:2014lca}.
For this mass, pure ULDM creates a slight enhancement at large scales, in addition to the expected suppression at smaller scales \textcolor{myyellow}{(dashed yellow)}.
The enhancement at larger scales is due to the modification of the background cosmology: before the onset of oscillations, $\phi$ acts like dark energy (DE) instead of a matter component, altering the time of matter-radiation equality and thereby affecting the relative peak heights of CMB~\cite{Marsh:2015xka,Hlozek:2014lca}.
Incorporating the effects of ULDM-induced VFCs impacts the CMB power spectra by increasing $Y_\mathrm{He}$ and modifying recombination \textcolor{myyellow}{(solid yellow)}.
Fig.~\ref{fig:xegz} indicates that the VFCs lead to advanced recombination and increased diffusion damping, resulting in the CMB peaks shifting to higher multipoles and an additional suppression of the CMB damping tail, respectively.
Moreover, the larger $Y_\mathrm{He}$ further suppresses the damping tail.

For $m_\phi = 10^{-31}$~eV in Fig.~\ref{fig:Cl_low_mass}, the situation is similar to the case of $m_\phi = 10^{-28}$~eV, except $\phi$ continues to behave as DE during and after recombination.
Therefore, the integrated Sachs–Wolfe effect is modified, affecting the CMB at large multipoles~\cite{Hlozek:2014lca,Baryakhtar:2024rky,Baryakhtar:2025uxs}(\textcolor{mygreen}{dashed green}).
The VFC at recombination is the dominant effect on the CMB; for the same amount of variation, the effect of $\delta Y_\mathrm{He}$ from BBN has a smaller impact~\cite{Seto:2022xgx} (dotted black).
Fig.~\ref{fig:xegz} shows that recombination is advanced substantially, and the reduced diffusing damping strongly enhances the CMB multipoles at small scales (\textcolor{mygreen}{solid green}).

Note that we do not incorporate the thermal mass term for our discussions in this subsection or in our low-mass analyses in Sec.~\ref{sec:mcmc-low}.
Accounting for the thermal mass is computationally expensive, and for our analysis choices, we expect our constraints to be largely unaffected by the thermal mass throughout most of the low-mass region.
However, depending on the choice of ULDM parameters, the thermal mass may modify the field evolution during not only BBN (see Fig.~\ref{fig:thmass}), but also recombination.
For example, for a fixed amount of variation, a very small value of $f_\phi$ would imply a correspondingly large value of the coupling $d_i^{(2)}$; thus, the field evolution would be dominated by the thermal mass for a prolonged period and produce a nontrivial VFC evolution during recombination.
If the thermal mass is relevant during recombination, we would need to properly account for the energy transfer between the ULDM and visible sectors; although the energy transfer at the background level is negligible (see Sec.~\ref{sec:model}), it may be important for the evolution of the ULDM perturbations.
We discuss the self-consistency of neglecting the thermal mass for our analysis results in Appendix~\ref{app:mcmctherm} and leave a proper treatment of the thermal mass to future work.

\section{Analysis and results}
\label{sec:mcmc}

From our discussion in Sec.~\ref{sec:cmb-combined}, the impact of ULDM with VFCs in the high-mass and low-mass regions are qualitatively different.
We perform separate analyses in these two regions to test ULDM with $\alpha$ and/or $m_e$ variations as a function of ULDM mass $m_\phi$.
We also investigate the viability of our model to fit CMB data with a larger value of $H_0$, compared to the $\Lambda$CDM best fit value.

\subsection{Analysis methodology}
We perform Markov chain Monte Carlo (MCMC) analyses for our quadratically coupled ULDM model using \texttt{MontePython}~\cite{Audren:2012wb,Brinckmann:2018cvx} to sample and \texttt{GetDist}~\cite{Lewis:2019xzd} to process our MCMC chains.
We employ the Metropolis–Hastings algorithm~\cite{Metropolis:1953am,Hastings:1970aa} and require a Gelman-Rubin convergence criterion~\cite{Gelman:1992zz} of $R-1 < 0.005$ for our high-mass analysis and $R-1 < 0.01$ for our low-mass analysis.
Achieving convergence for the low-mass analysis takes longer due to the larger number of parameters, as well as the presence of parameter degeneracies.

We use wide flat priors for the 6 standard $\Lambda$CDM parameters: $\{ h, \omega_b, \omega_{\rm CDM}, \ln (10^{10} A_s), n_s, \tau_{\rm reio} \}$.
To analyze the high-mass region, we vary the 6 $\Lambda$CDM parameters, as well as $\delta Y_{\rm He} \in [0,\infty)$ with a flat prior.
To analyze the low-mass region, we vary the $\Lambda$CDM parameters, except we replace the CDM energy density $\omega_{\rm CDM}$ with the total dark matter relic density $\Omega_{\rm DM}$, for which we also use a wide flat prior.
Additionally, we vary the ULDM parameters $\log_{10}f_\phi \in [-3,0]$, $\log_{10}(m_\phi/{\rm eV}) \in [-33,-23]$, and $\log_{10}(d_i^{(2)}f_\phi) \in [-4,2]$ with $i=\{\alpha, m_e\}$ and flat priors.

The datasets we use for our analyses are as follows:
\begin{itemize}
\item {\it Planck}: We use the {\it Planck} 2018 low-$\ell$ TT and low-$\ell$ EE $(2\leq\ell\leq 29)$ likelihoods. For high-$\ell$, we use the TTTEEE likelihood ($30 \leq \ell \leq 2508$ for TT and $30 \leq \ell \leq 1996$ for EE)~\cite{Planck:2019nip}.
  We also include the {\it Planck} lensing likelihood~$(8< L \lesssim 400)$~\cite{Planck:2018lbu}.
\item South Pole Telescope (SPT): We use the SPT-3G Y1 data release,\footnote{\url{https://github.com/SouthPoleTelescope/spt3g_y1_dist}} which includes TT $(750<\ell\lesssim 3000)$ and TEEE 
$(300<\ell\lesssim 3000)$~\cite{SPT-3G:2021wgf,SPT-3G:2022hvq}.
\item Baryon acoustic oscillation (BAO) measurements: We use the 6DF Galaxy Survey~\cite{Beutler:2011hx}, SDSS-DR7 MGS~\cite{Ross:2014qpa}, and BOSS DR12 (including $f\sigma_8$)~\cite{BOSS:2016wmc}.
\end{itemize}
For both \textit{Planck} and SPT, we vary all of the nuisance parameters included with the likelihoods in \texttt{MontePython}.
Note that we do not account for nonlinear corrections to compute the CMB power spectra; such corrections may be important for small-scale CMB data and should account for any power suppression from ULDM~\cite{Trendafilova:2025dce}.

\subsection{Constraints for high-mass region}
\label{sec:mcmc-high}

In the high-mass region $(m_\phi \gtrsim 10^{-23}~{\rm eV})$, the only non-negligible effect of the VFCs on the CMB is through modification of $Y_\mathrm{He}$ from BBN, and ULDM evolves as CDM.
Therefore, our analysis can be simplified to a fit of $\Lambda$CDM with an additional parameter $\delta Y_{\rm He}$ to account for the change in the helium-4 abundance.
We can then map the resulting constraint on $\delta Y_{\rm He}$ into a constraint on $m_\phi$ and the couplings, $d_\alpha^{(2)}$ and/or $d_{m_e}^{(2)}$, for a given ULDM fraction $f_\phi$.
We consider only positive ULDM couplings $d_i^{(2)}$, and in the parameter space of interest, we always have $\delta Y_{\rm He} \geq 0$.
Therefore, we restrict the prior on $\delta Y_{\rm He}$ to be non-negative and use
\begin{equation}
    \label{eq:DYp_var}
    Y_\mathrm{He} = Y_\mathrm{He}^{\rm SM} (1 + \delta Y_\mathrm{He}) \;,
\end{equation}
where $Y_\mathrm{He}^{\rm SM}$ is the standard BBN prediction for a given $\omega_b$ and $N_{\rm eff}$ computed by \texttt{CLASS}.
We fix $N_{\rm eff} = 3.044$ to its SM prediction~\cite{Akita:2020szl, Froustey:2020mcq, Bennett:2020zkv}.\footnote{Variation of the electron mass or $\alpha$
can change the evolution of the effective number of relativistic degrees of freedom $(g_\star)$ around the freeze-out of $e^+ e^-$ annihilation. The entropy leakage between photons and neutrinos around neutrino decoupling may be affected and slightly modify the calculation of $N_{\rm eff}$. We ignore this possibility in our work.}

We obtain the following upper limit on $\delta Y_{\rm He}$ at the 95\% confidence level (C.L.):
\begin{equation}
    \label{eq:DYHe constraints}
    \delta Y_{\rm He} ~(95\%~{\rm C.L.}) <
    \begin{cases}
        8.4\% & \textit{Planck} \\
        6.7\% & \textit{Planck} + \textrm{BAO} + \textrm{SPT} \;.
    \end{cases}
\end{equation}
Primordial $Y_\mathrm{He}$ can also be inferred from metal-poor extragalactic regions, and the recommended value from the Particle Data Group (PDG) is $Y_\mathrm{He} = 0.245 \pm 0.003$~\cite{ParticleDataGroup:2024cfk}, corresponding to $\delta Y_\mathrm{He} < 2.4\%$ at $2\sigma$.
For a given mass $m_\phi$ and fraction $f_\phi$, we can solve for the evolution of $\phi$ and use Eq.~\eqref{eq:DY_p} to determine the values of the ULDM couplings that yield a particular value of $\delta Y_\mathrm{He}$.
This procedure allows us to translate our CMB bound or the PDG bound to a corresponding limit on the ULDM couplings.

\begin{figure}
    \centering
    \includegraphics[width=0.49\linewidth]{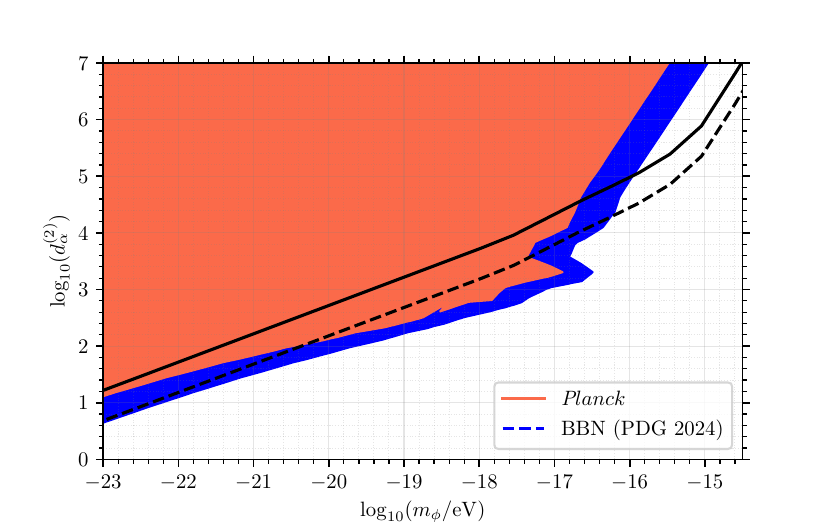}
    \includegraphics[width = 0.49\linewidth]{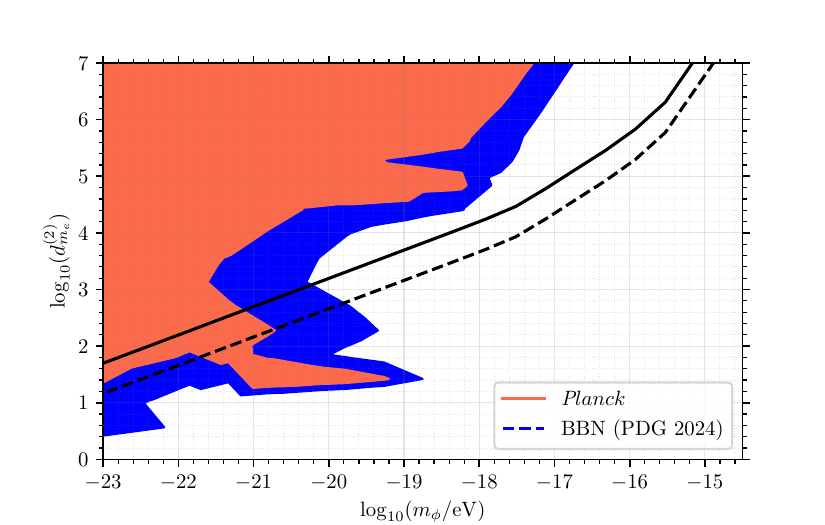}
    \caption{95\% C.L.~constraints in the ULDM high-mass region for the ULDM couplings $d_\alpha^{(2)}$ (left) and $d_{m_e}^{(2)}$ (right), which induce a change in $Y_\mathrm{He}$ from the variation of $\alpha$ and $m_e$, respectively.
      We consider one nonzero coupling at a time, and we fix the ULDM fraction to $f_\phi = 1$.
      The constraints labeled as {\it Planck} and BBN correspond to the 95\% C.L.~limits on $Y_\mathrm{He}$ from our simplified analysis [see Eq.~\eqref{eq:DYHe constraints}] and from the PDG~\cite{ParticleDataGroup:2024cfk}, respectively.
      The shaded exclusion regions take into account $m_{\rm th}$ effects, while the black lines show the corresponding limit obtained from neglecting thermal mass effects.
      The constraints obtained from neglecting the thermal mass can actually be considered constraints on the combination $f_\phi d_i^{(2)}$.
      This degeneracy is broken when including the thermal mass, so the shaded constraints hold specifically for $f_\phi = 1$.
      }
    \label{fig:BBN_constraint}
\end{figure}

Fig.~\ref{fig:BBN_constraint} shows the exclusion regions for variations in $\alpha$ (with no $m_e$ variations) and $m_e$ (with no $\alpha$ variations) from our CMB result in Eq.~\eqref{eq:DYHe constraints} and the PDG value for $Y_{\rm He}$.
The black solid and dashed curves correspond to the constraints derived from neglecting the effect of the thermal mass, while the shaded regions show the excluded parameter space when properly accounting for the thermal mass.
The impact of the thermal mass decreases for smaller $m_\phi$, so the constraints with and without the thermal mass coincide at small $m_\phi$, where the thermal mass effects are negligible.
We provide further details in Appendix~\ref{app:mcmctherm}. The constraints of the couplings for the PDG bound on $\delta Y_{\rm He}$ are consistent with the ones derived in Ref.~\cite{Bouley:2022eer} for both types of variation. 

The constraints obtained when neglecting the thermal mass have minimal features: they scale as $\propto m_\phi^{-1/2}$ for $m_\phi \lesssim 10^{-15}{\rm eV}$ and $\propto m_\phi^2$ for $m_\phi \gtrsim 10^{-15}{\rm eV}$.
Furthermore, note that the variations in Eqs.~\eqref{eq:mevar} and \eqref{eq:alphavar} depend on the combination $\phi^2 d_i^{(2)}$.
Since $\rho_\phi \propto m_\phi^2\phi^2$, the effect of the VFCs is sensitive to the combination $f_\phi d_i^{(2)}$.
Therefore, the constraints without thermal mass effects in Fig.~\ref{fig:BBN_constraint}, for which we fix $f_\phi = 1$, can actually be considered constraints on $f_\phi d_i^{(2)}$.
This degeneracy allows for an extraction of the limit on $d_i^{(2)}$ for $f_\phi < 1$.

The thermal masses in Eqs.~\eqref{eq:thmassme} and \eqref{eq:thmassgamma} depend on the couplings, unaccompanied by a factor of $\phi^2$, so the thermal mass breaks this degeneracy between $d_i^{(2)}$ and $f_\phi$.
As shown in Fig.~\ref{fig:thmass}, the evolution of $\phi$ is much more nontrivial when accounting for the effects of the thermal mass, and the resulting constraints in Fig.~\ref{fig:BBN_constraint} exhibit jagged features, especially for the variation of $m_e$.
Additionally, these constraints are valid for our choice of $f_\phi = 1$ and cannot be rescaled to find the limit on $d_i^{(2)}$ for $f_\phi < 1$.

\begin{figure}
    \centering
    \includegraphics[width=0.6\linewidth]{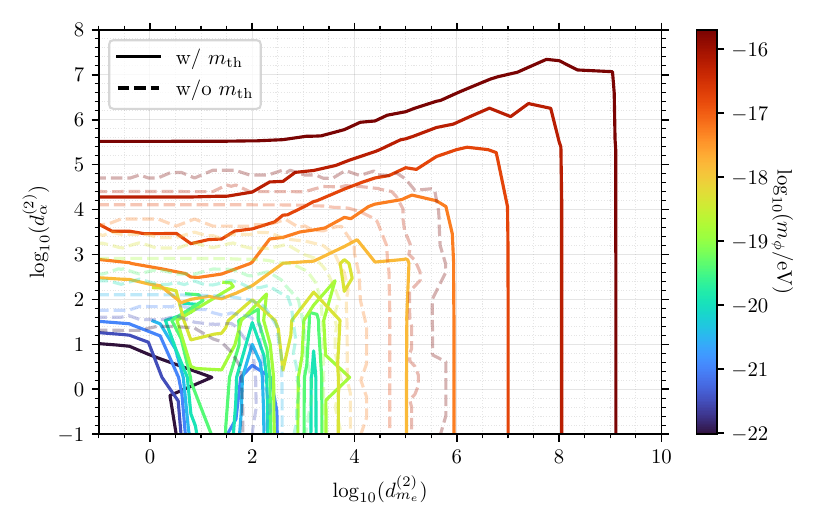}
    \caption{$95\%$ constraints in the ULDM high-mass region for the ULDM couplings $d_\alpha^{(2)}$ and $d_{m_e}^{(2)}$, which are allowed to vary simultaneously, using the constraints on $Y_{\rm He}$ from the PDG~\cite{ParticleDataGroup:2024cfk}.
      Region above the corresponding line for a given mass is excluded.
      The solid and dashed lines correspond to the constraints obtained from including and ignoring the thermal mass, respectively.
      }
    \label{fig:HM_em_vary}
\end{figure}

Fig.~\ref{fig:HM_em_vary} shows the constraints obtained from varying $d_\alpha^{(2)}$ and $d_{m_e}^{(2)}$ simultaneously with (solid lines) and without (dashed) the effects of the thermal mass.
We only show the constraints using the PDG bound, since it is stronger than our CMB bound on $\delta Y_{\rm He}$.
For a given mass, couplings higher than the respective line are excluded.
Accounting for the thermal mass weakens the constraint on either coupling for higher masses and strengthens it at lower masses, which can also be seen in Fig.~\ref{fig:BBN_constraint}.

\subsection{Constraints for low-mass region}
\label{sec:mcmc-low}

The combined effect of the ULDM field and the VFCs it induces is relevant for the CMB in the low-mass regime $(m_\phi \lesssim 10^{-23}~{\rm eV})$.
The background and perturbation evolution of ULDM does not mimic CDM, and the VFCs increase $Y_\mathrm{He}$ and affect recombination.
As noted in Sec.~\ref{sec:cmb-combined}, we neglect the effects of the thermal mass and discuss this approximation in further detail in Appendix~\ref{app:mcmctherm}.

To study the rich phenomenology of ULDM and VFCs, we perform MCMC analyses that vary $f_\phi$, $m_\phi$, and $d_i^{(2)}$ in addition to the standard $\Lambda$CDM.
Since we neglect the thermal mass in this regime, we vary $f_\phi d_i^{(2)}$ instead of $d_i^{(2)}$, as the amount of variation is proportional to the combination of these parameters (see Sec.~\ref{sec:mcmc-high}).
To ensure the convergence of our sampling across a wide range of $m_\phi$, we perform three MCMC analyses that correspond to three mass subdivisions in which the limits on $f_\phi$ and $f_\phi d_i^{(2)}$ differ appreciably: $m_{\phi}^{\rm (low)} \in [10^{-33},10^{-31}]~{\rm eV}$, $m_{\phi}^{\rm (mid)} \in [10^{-31},10^{-26}]~{\rm eV}$, and $m_{\phi}^{\rm (high)} \in [10^{-26},10^{-23}]~{\rm eV}$.
We have verified that our results using these mass subdivisions are consistent with our main analysis that varies $m_\phi$ across the full mass range of interest.

\begin{figure}
    \centering
    \includegraphics[width=0.49\textwidth]{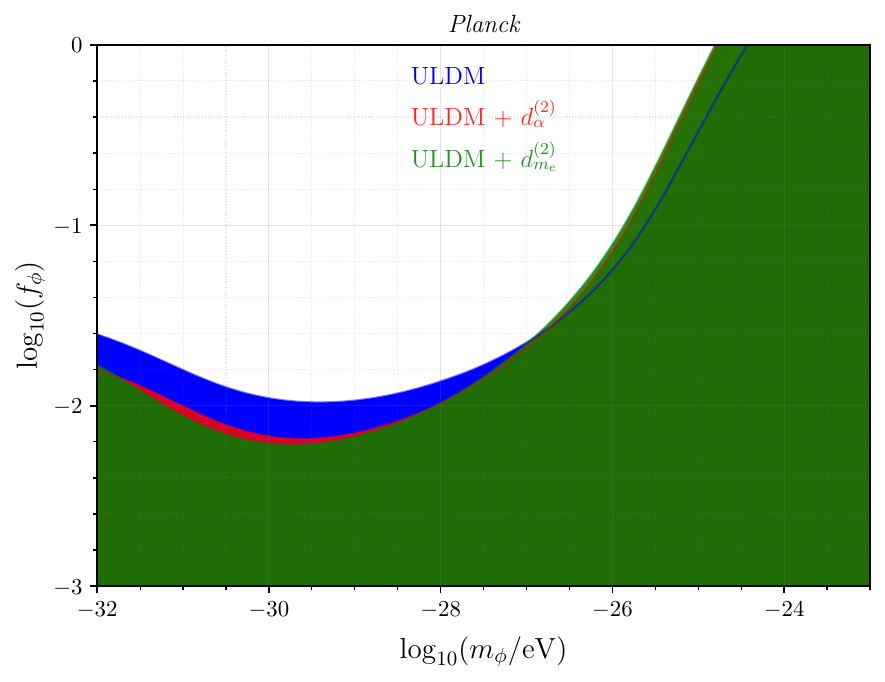}
    \includegraphics[width=0.49\textwidth]{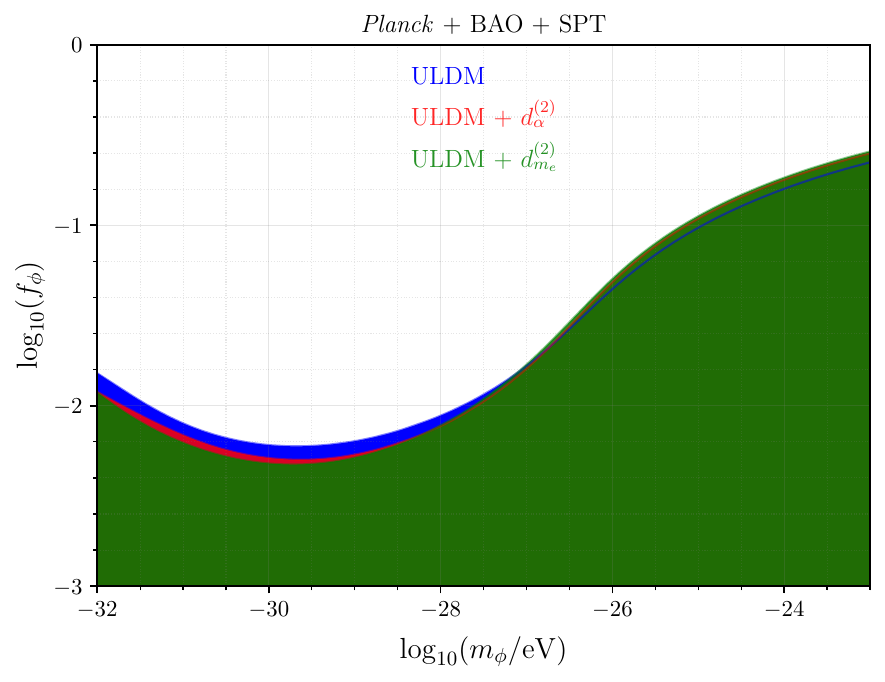}
    \caption{Marginalized $95\%$ C.L.~posteriors in the plane of $f_\phi$ vs.~$m_\phi$ for \textit{Planck} (left) and \textit{Planck} + BAO + SPT (right).
      We analyze the case of pure ULDM (blue), ULDM with nonzero coupling $d_\alpha^{(2)}$ (red), and ULDM with nonzero coupling $d_{m_e}^{(2)}$ (blue).
      The effects of the VFCs tighten the constraint on $f_\phi$ at low masses.
      Note that the thermal mass is neglected in the VFC analyses.
      }
    \label{fig:2D_frac}
\end{figure}

Fig.~\ref{fig:2D_frac} shows the constraint on $f_\phi$ as a function of $m_\phi$ in terms of the 2D marginalized 95\% C.L.~posteriors, obtained from analyzing \textit{Planck} (left) and \textit{Planck} + BAO + SPT (right).
We show the results for three analyses: pure ULDM (blue), ULDM with $d_\alpha^{(2)} \neq 0$ (red), and ULDM with $d_{m_e}^{(2)} \neq 0$ (blue).
Our pure ULDM results are consistent with those reported in Refs.~\cite{Lague:2021frh,Rogers:2023ezo}.
For $m_\phi \lesssim \textrm{few}\times 10^{-27}$~eV, we find that the effects of VFCs are important and improve the constraint on $f_\phi$ over pure ULDM when fitting to \textit{Planck}.
Although improvement is smaller for \textit{Planck} + BAO + SPT, the additional datasets provide more constraining power on $f_\phi$ across all masses.
In particular, at high masses of $m_\phi \gtrsim 10^{-25}$~eV, the joint constraint is $f_\phi \lesssim 0.3$.

\begin{figure}
    \centering
    \includegraphics[width=0.49\linewidth]{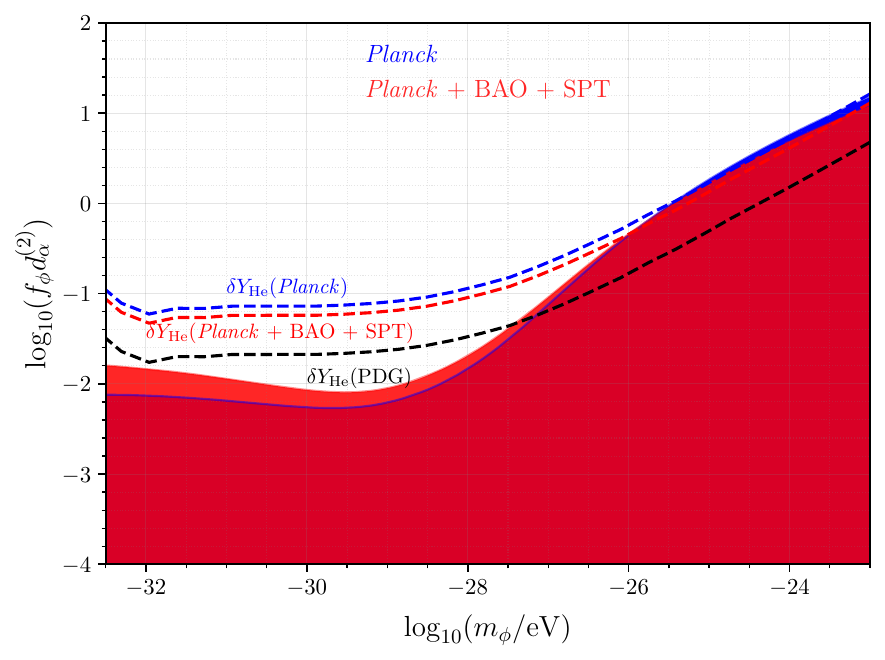}
    \includegraphics[width=0.49\linewidth]{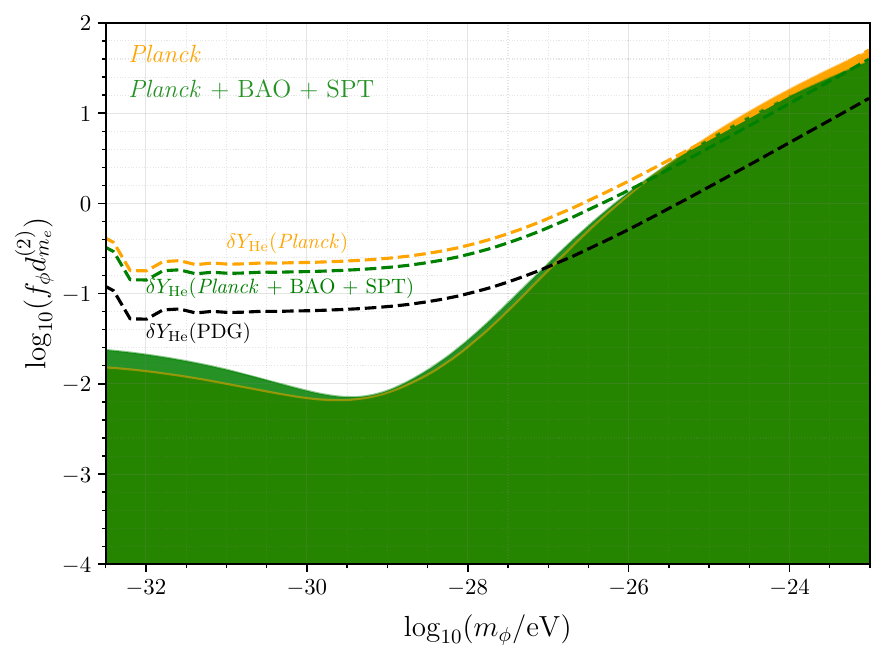}
    \caption{Marginalized 95\% C.L.\ posteriors in the plane of the combined parameter $f_\phi d_i^{(2)}$ vs.~$m_\phi$ for \textit{Planck} and for \textit{Planck} + BAO + SPT.
      We analyze $\alpha$-only variations (left) and $m_e$-only variations (right) separately.
      Dashed lines correspond to the 95\% C.L.\ upper limit on $\delta Y_{\rm He}$ (see Sec.~\ref{sec:mcmc-high}) from the PDG~\cite{ParticleDataGroup:2024cfk} and from $\Lambda$CDM+$\delta Y_{\rm He}$ analyses of the indicated datasets.
      Towards the upper end of the mass range, the colored dashed lines approach the solid contours, indicating that ULDM behaves as CDM and the effect of VFCs is relevant only at BBN for $m_\phi \gtrsim 10^{-26}$ eV.
      The full analysis of ULDM with SM couplings provides stronger constraints on $f_\phi d_i^{(2)}$, compared to the simpler $\delta Y_{\rm He}$ analysis, for $m_\phi \lesssim 10^{-26}$ eV, where the recombination modifications are relevant.
      CMB data provide a stronger constraint than PDG $\delta Y_{\rm He}$ considerations alone for $m_\phi \lesssim 10^{-28}$ eV.
      Note that the thermal mass is neglected for all analysis results shown.
      }
    \label{fig:2D_constraint}
\end{figure}

Fig.~\ref{fig:2D_constraint} shows the constraint on $f_\phi d_i^{(2)}$ as a function of $m_\phi$ in terms of the 2D marginalized 95\% C.L.~posteriors, obtained from analyzing \textit{Planck} and \textit{Planck} + BAO + SPT.
We consider ULDM with $d_\alpha^{(2)} \neq 0$ (left) and ULDM with $d_{m_e}^{(2)} \neq 0$ (right).
The dashed lines show the 95\% C.L.\ upper limit from $\delta Y_\mathrm{He}$ considerations only and are applicable for the analysis of the high-mass region described in Sec.~\ref{sec:mcmc-high}.
The \textit{Planck} and PDG lines are the continuation of the limits (neglecting thermal mass effects) in Fig.~\ref{fig:BBN_constraint} to lower values of $m_\phi$.
Analogous to the \textit{Planck} limit presented in Sec.~\ref{sec:mcmc-high}, the \textit{Planck} + BAO + SPT limit of $\delta Y_{\rm He} < 6.7\%$ at 95\% C.L.\ is obtained from the $\Lambda$CDM+$\delta Y_{\rm He}$ analysis of all three datasets.

Our full analyses of ULDM with SM couplings in the low-mass region account for the evolution of ULDM and the VFC that both change $Y_{\rm He}$ and modify recombination.
In the regime $m_\phi \gtrsim 10^{-26}~{\rm eV}$, the MCMC constraints approach the corresponding constraints from $\delta Y_\mathrm{He}$ only.
This agreement indicates that ULDM behaves as CDM for these masses and that the only effect of the VFC is through the change in $Y_\mathrm{He}$ at BBN, while the modification at recombination is negligible.
For masses $\textrm{few}\times 10^{-28}~{\rm eV} \lesssim m_\phi \lesssim 10^{-26}~{\rm eV}$, the full analysis of coupled ULDM yields stronger constraints over the $Y_\mathrm{He}$ analysis: both $f_\phi$ and $d_i^{(2)}$ individually need to be smaller due to the otherwise large suppression of the CMB damping tail from the modifications to recombination and the ULDM field itself, as demonstrated in Fig.~\ref{fig:Cl_low_mass}.
For low masses of $m_\phi \lesssim \textrm{few}\times 10^{-27}~{\rm eV}$, there is a significant improvement of the constraint compared to the corresponding $\delta Y_\mathrm{He}$ constraint.
Moreover, the constraint from the full analysis of coupled ULDM in this regime surpasses the PDG $\delta Y_\mathrm{He}$ constraint.

\subsection{Implications for Hubble}
\label{sec:mcmc-hubble}

\begin{figure}
    \centering
    \includegraphics[width=0.49\linewidth]{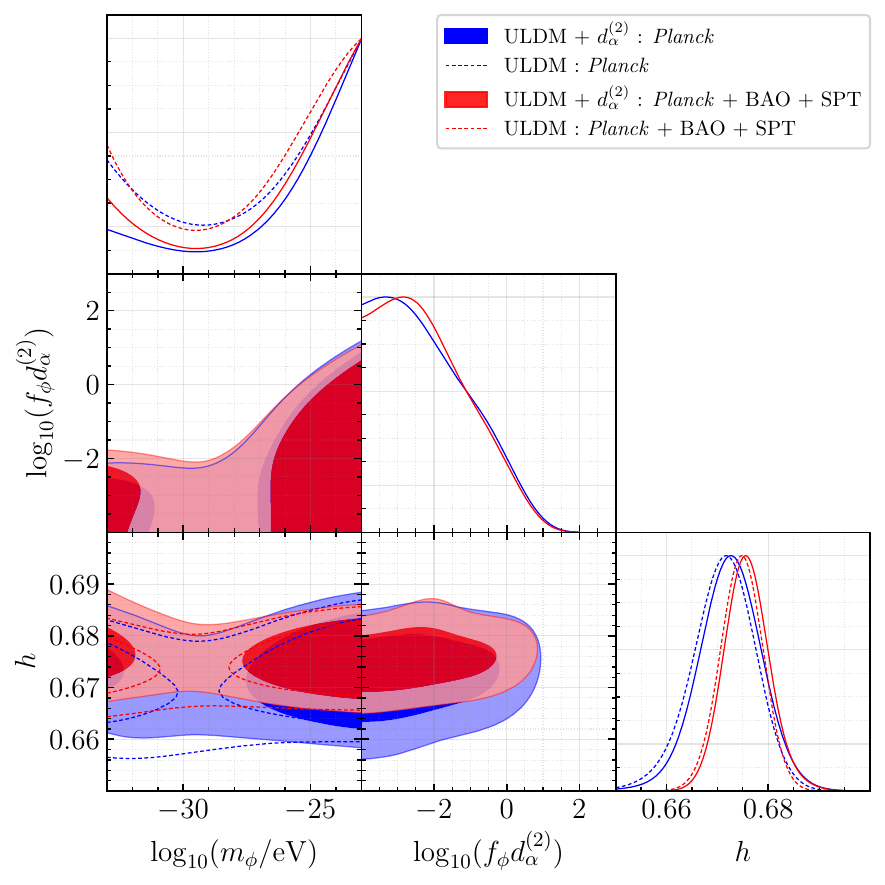}
    \includegraphics[width=0.49\linewidth]{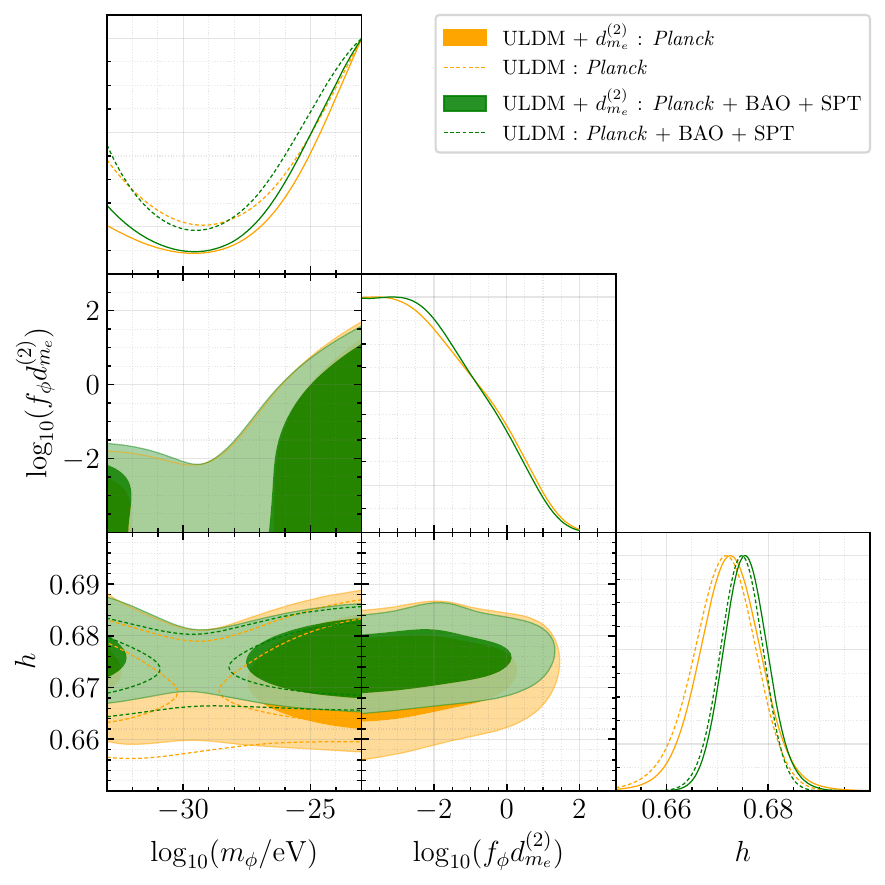}
    \caption{Triangle plot with the 68\% and 95\% C.L.~contours for the marginalized posteriors of the ULDM mass $m_\phi$, the VFC parameter $f_\phi d^{(2)}_{i}$, the reduced Hubble constant $h$, and the derived parameter $\delta Y_{\rm He}$ 
    for $\alpha$ variation (left) and $m_e$ variation (right).
      We analyze both the pure ULDM scenario (dotted) and the ULDM-induced VFC scenario (solid) with \textit{Planck} and with \textit{Planck} + BAO + SPT.
      }
    \label{fig:tri_h0}
\end{figure}

In this section, we explore how oscillatory VFCs could impact the inferred value of the Hubble constant $H_0$.
In Fig.~\ref{fig:tri_h0}, we show the triangle plots of the 1D and 2D marginalized posteriors for $m_\phi$, $f_\phi d_i^{(2)}$ and $h~ (\equiv H_0/(100{\rm~ km/s/Mpc}))$ from our analyses of the \textit{Planck} and \textit{Planck} + BAO + SPT datasets for ULDM with $\alpha$ variation (left) and ULDM with $m_e$ variation (right).
We also provide the outlines of the posterior distributions obtained from our pure ULDM analyses.
The mean value of $h$ exhibits a negligible difference compared to pure ULDM.

The 2D $h$ vs.~$m_\phi$ contours show the impact of VFC on the Hubble constant for different $m_\phi$ values.
In the $m_\phi \gtrsim 10^{-27}~{\rm eV}$ region, VFC modifications only increase $\delta Y_\mathrm{He}$ which has negligible correlation with $h$. Therefore, in that region, the $h$ contours remain almost unchanged compared to the pure ULDM analysis.
Fig.~\ref{fig:tri_h0} indicates that the oscillatory features in $m_e$ induced for $m_\phi \approx 10^{-28}~{\rm eV}$ (see Fig.~\ref{fig:Dxe}) have no appreciable impact on $H_0$, in contrast to the increase of $H_0$ in Ref.~\cite{Lee:2022gzh} due to oscillatory-like variations of $m_e$, with the amplitude of variations increasing as recombination progresses.

Taking only the portion of our MCMC chains that lie in the mass range $m_\phi < 10^{-29}~{\rm eV}$, which is associated with a constant VFC at recombination, we find a small increase in the mean value of $h$, compared to pure ULDM. This higher value can be traced to a delay of recombination in the presence of VFCs~\cite{Hart:2019dxi,Sekiguchi:2020teg,toda_constraints_2025,Hart:2021kad,Seto:2022xgx,Lee:2022gzh,Schoneberg:2021qvd,Zhang:2022ujw,Schoneberg:2024ynd,Baryakhtar:2024rky,Baryakhtar:2025uxs,ACT:2025tim}, among other effects from $\delta Y_{\rm He}$ and ULDM perturbations, but the increase is not statistically significant.
Additional plots and tables depicting the constraints of all $\Lambda$CDM and VFC parameters are shown in Appendix~\ref{app:traingle_plots}.

\section{Conclusion}
\label{sec:conclusion}

In this work, we have studied the cosmological implications of a well-motivated scenario in which a variation of fundamental constants is induced by a scalar field coupled to the SM.
The evolution of the scalar field, which also behaves as ULDM, induces redshift-dependent VFCs.
We study these effects for $\alpha$ and $m_e$ variations through their modifications to $Y_\mathrm{He}$ during BBN and to recombination dynamics during the CMB era.
We explore how oscillatory VFCs during recombination alter the CMB power spectra, and we study the interplay between the impact on BBN vs.~the CMB.

Using \textit{Planck} 2018 data, as well as SPT-3G and BAO measurements, we derive constraints across a range of ULDM masses relevant for the CMB. The analyses can be organized in terms of two mass ranges. For the ``low-mass'' region where $m_\phi \lesssim 10^{-23}$~eV,  
our analyses show that the presence of VFCs strengthens the constraint on the fraction of ULDM, compared to the constraint from a pure ULDM analysis with no SM couplings.
We find that a large $\phi$-induced VFC during recombination is strongly constrained by CMB data, much stronger than bounds arising from the inferred value of $Y_\mathrm{He}$ from astrophysical measurements.
For the ``high-mass'' region where $m_\phi \gtrsim 10^{-23}$~eV, we are able to simplify the analysis and perform a self-consistent study that incorporates the thermal mass for $\phi$, generated by the ULDM coupling to the thermal SM bath. We find bounds that are consistent with the BBN analysis of Ref.~\cite{Bouley:2022eer}.

We consider only positive couplings $d_i^{(2)} > 0$, which induce a positive shift in the value of the fundamental constants.
The science case for negative couplings is also intriguing, in which case the values of the constants are smaller in the early Universe compared to their canonical values.
Negative coupling corresponds to a negative thermal mass for the scalar field, which can create tachyonic instabilities~\cite{Sibiryakov:2020eir,Bouley:2022eer} that can give rise to an exponential enhancement of the scalar field energy density.
The phenomenology of a negative coupling is beyond the scope of this paper, and we leave it to future work.
We also consider only the $\phi^2$ coupling to the SM, which is less constrained than the linear $\phi$ coupling from fifth-force experiments~\cite{Adelberger:2003zx}.

Finally, we focus on ULDM couplings that alter the electron mass and fine structure constant.
The additional couplings in Eq.~\eqref{eq:Ldamour} have been studied in the context of BBN~\cite{Bouley:2022eer}.
We expect these couplings to have interesting implications for cosmology, beyond the modifications to $Y_{\rm He}$ during BBN, and we will explore this possibility in upcoming work.

\begin{acknowledgments}
We thank Masha Baryakhtar, Jens Chluba, Tanvi Karwal, Olivier Simon, Zachary J. Weiner, and Yue Zhao for useful discussions. KB and SG acknowledge support from the National Science Foundation (NSF) under Grant No.~PHY-2413016 and acknowledge the \textit{Dark Matter Theory, Simulation, and Analysis in the Era of Large Surveys} workshop and the Kavli Institute for Theoretical Physics for the hospitality and support under NSF grant No.~PHY-2309135 during the intermediate stages of this work.  T-TY is supported in part by NSF CAREER grant PHY-1944826. T-TY thanks the hospitality of the
Università degli Studi di Padova and the CERN Theory group where portions of this work were
completed. 
\end{acknowledgments}

\appendix

\section{\texttt{scalarCLASS}}
\label{app:classbbn}
\begin{figure}
    \centering
    \includegraphics[width=0.7\linewidth]{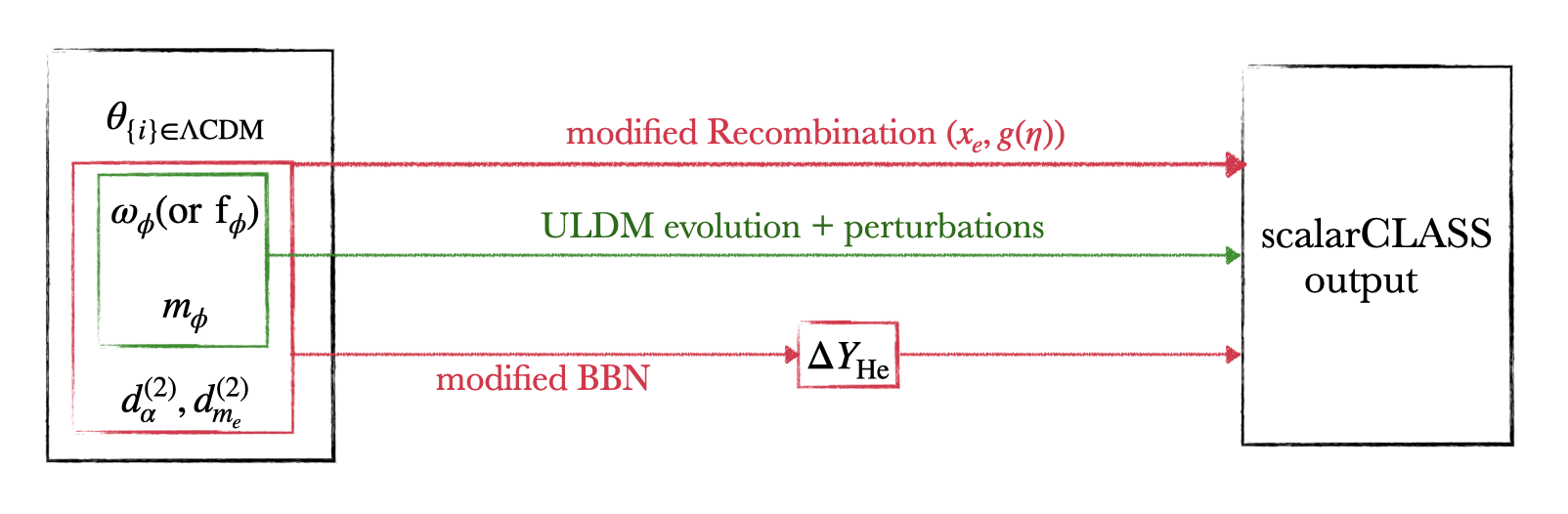}
    \caption{Schematic diagram of \texttt{scalarCLASS}, our modified version of \texttt{CLASS} that computes the effects of $\phi$-induced VFCs on the CMB and matter power spectrum.
      The ULDM background and perturbation evolutions implemented in the code are based on \texttt{AxiCLASS}.
      The VFCs alter the predicted value of $Y_{\rm He}$ from BBN, which is computed within \texttt{scalarCLASS}.
      They also modify recombination via changes to the free-electron fraction $x_e$ and the visibility function $g({\eta})$.
      }
    \label{fig:flow_chart}
\end{figure}

In this section, we detail the main features of \texttt{scalarCLASS}, which we developed for this work to study ULDM-induced VFCs.
The code is based on \texttt{AxiCLASS}~\cite{Smith:2019ihp,Poulin:2018dzj,Poulin:2018cxd,Murgia:2020ryi}, which is an extension of the \texttt{CLASS}~\cite{Lesgourgues:2011re} code that incorporates the Boltzmann hierarchy of a cosmological scalar field for a variety of potentials relevant for, e.g., axion-like particles or early dark energy.
We implement the potential $V = {1 \over 2} m_{\rm eff}^2\phi^2$ for the ULDM field $\phi$.
A schematic outline of the code is shown in Fig.~\ref{fig:flow_chart}.

\subsection{Derivation of $Y_{\rm He}$}

One of the main features of this code is that it automatically computes the BBN era modifications of $Y_{\rm He}$ induced by VFCs.
We solve the system of coupled differential equations in Eqs.~\eqref{eq:rateeqn}, \eqref{eq:neutron-decay-abundance}, and \eqref{eq:Drateeq} to compute $\Delta X_n/X_n$ at $a = a_w$, i.e, the weak-decoupling time.
This is one of the primary contributions to $\Delta Y_{\rm He}/Y_{\rm He}$ at BBN, which is the first term on the right-hand side of Eq.~\eqref{eq:DY_p}.
The other contribution to $\Delta Y_{\rm He}/Y_{\rm He}$ comes from the decay of neutrons after weak decoupling until the onset of BBN, represented by the second term on the right-hand side of Eq.~\eqref{eq:DY_p}.
We implement these computations in the \texttt{background} module of \texttt{scalarCLASS}. The \texttt{background} module in \texttt{CLASS} computes a set of integrated (e.g., sound horizon) and non-integrated (e.g., energy densities) quantities as a function of scale factor and stores them in numerical tables from some initial scale factor $(a_{\rm ini})$ through today $(a = 1)$.
We use this built-in structure to implement the BBN calculation.
We break down the neutron decay contribution in the following manner:
\begin{align}
    \label{eq:DYHe_2nd_term}
    \int_{a_w}^{a_{\rm BBN}} f(\phi(a),a)d\log a &= \int_{a_{\rm ini}}^{a_{\rm BBN}} f(\phi(a),a)d\log a -  \int_{a_{\rm ini}}^{a_W} f(\phi(a),a)d\log a,\\
    f(\phi(a),a) &= \dfrac{\Gamma_n}{H}\left[5{\Delta m_e \over m_e} - \dfrac{m_{np}}{m_e}\dfrac{P'}{P}\left(\dfrac{\Delta m_e}{m_e} - \dfrac{\Delta m_{np}}{m_{np}}\right)\right]\;.
\end{align}
In \texttt{scalarCLASS}, we numerically solve for the integrated quantities $\Delta X_n/X_n$ and $\int_{a_{\rm ini}}^{a}f(\phi(a),a)d\log a$, and we store them in a background table as a function of scale factor.
Using the values of these quantities at $a_{w}$ and $a_{\rm BBN}$, we compute $\Delta Y_{\rm He}/Y_{\rm He}$ as given in Eq.~\eqref{eq:DY_p}.
For our calculations, we use $a_{\rm ini} = 10^{-15}$. 
At the initial time $a_{\rm ini}$, neutrons and protons are in equilibrium.
We set $X_n(a_{\rm ini}) = X_n^{\rm eq}$ and $\Delta X_n(a_{\rm ini}) = \Delta X_n^{\rm eq}$, which are given in Eqs.~\eqref{eq:eqabandance} and \eqref{eq:DXneq}, respectively.

\subsection{Implementation of $g_\ast$}

Various rates governing the computation of $\delta Y_{\rm He}$ depend on the temperature of the photon bath, which can be calculated as a function of the scale factor.
The standard way of computing temperature in \texttt{CLASS} is
\begin{equation}
    \label{eq:T_class}
    T(a) = T_0/a\;,
\end{equation}
where $T_0$ is the current CMB temperature.
This is, however, inadequate for our purposes, since it cannot be extended to higher temperatures beyond BBN, when the effective number of relativistic degrees of freedom in the bath $g_\ast (a)$ differs from today.
As the thermalized particles of the SM become nonrelativistic as the Universe cools, their entropy is transferred to the bath, thereby causing the bath temperature to decrease more slowly than in Eq.~\eqref{eq:T_class}.
In \texttt{scalarCLASS}, we implement the following definition of temperature:
\begin{equation}
    \label{eq:T_scalarCLASS}
    T(a) = T_0 \left(g_\ast(a) \over g_\ast(0)\right)^{1/3}\;.
\end{equation}
We use $g_\ast(T)$ from Ref.~\cite{Planck:2018vyg} and convert it to $g_\ast(a)$ via Eq.~\eqref{eq:T_scalarCLASS}.
We encode $g_\ast(a)$ through an interpolation function in \texttt{scalarCLASS}.

\subsection{Computation of thermal mass $m_{\rm th}$}

The coupling of $\phi$ to the SM induces a thermal-mass contribution for $\phi$.
In accordance with Eq.~\eqref{eq:eqm}, the equation of motion for $\phi$ is
\begin{equation}
  \ddot{\phi} + 2aH\dot{\phi} + a^2m_{\rm eff}^2\phi = 0\;,
\end{equation}
where 
\begin{align}
    m_{\rm eff}^2(T) &= m_\phi^2 + m_{\rm th}^2(T)\\
    m_{{\rm th},m_e}^2 (T) &= \dfrac{2\pi d_{m_e}^{(2)}}{M_{\rm Pl}^2}\dfrac{4m_e^2}{\pi^2}T^2\int_{m_e/T}^\infty dx \dfrac{\sqrt{x^2 -(m_e/T)^2}}{e^x + 1}\;, \\
  m_{{\rm th},\alpha}^2 (T) &\simeq \dfrac{2\pi d_{\alpha}^{(2)}}{M_{\rm Pl}^2} \dfrac{\alpha}{4\pi}\dfrac{\pi^2}{3}T^4\;,
\end{align}
following Eqs.~\eqref{eq:thmassme} and \eqref{eq:thmassgamma}.
The implementation of $m_{{\rm th},\alpha}^2$ in the code is straightforward.
For $m_{{\rm th},m_e}^2 (T)$, we need to compute the integral in Eq.~\eqref{eq:thmassme}.
For computational efficiency, we compute the integral for a range of $m_e/T$ values and store them in a table; we use an interpolation routine to incorporate them into \texttt{scalarCLASS}.

\subsection{Implementation of $\phi$-induced VFCs}
Simple parameterizations of $\alpha$ and $m_e$ variations exist in the current version \texttt{CLASS}, following Ref.~\cite{Hart:2017ndk}.
We adapt these parameterizations to incorporate ULDM-induced VFCs in \texttt{scalarCLASS}.
The built-in formalism in \texttt{CLASS} only considers \emph{constant} modifications of fundamental constants (with a cutoff in redshift, after which the variation is set to zero).
By modifying appropriate modules, we allow the variation to be \emph{redshift-dependent}.
The amount of variation at a given redshift is computed using Eqs.~\eqref{eq:mevar} and \eqref{eq:alphavar}:
\begin{align}
  \dfrac{\Delta m_e(z)}{m_e} &= 2\pi d_{m_e}^{(2)}\dfrac{\phi^2(z)}{M_{\rm pl}^2}\;, \\
  \dfrac{\Delta\alpha(z)}{\alpha} &= 2\pi d_\alpha^{(2)}\dfrac{\phi^2(z)}{M_{\rm pl}^2}\;.
\end{align}
Note that in our formalism, $\phi$ undergoes oscillations; thus, the values of the constants vary rapidly at late times.
For most cases, scaling the rates of recombination processes with the modified values of the constants is sufficient to include the effects of redshift-dependent VFCs.
However, the timescale of the change in the values of constants can be fast compared to the lifetime of atomic transitions relevant for recombination, which may necessitate a more careful treatment of VFCs.

Fortunately, in the region where the VFC is highly oscillatory, the magnitude of the change is also very small since $\phi \sim a^{-3/2}$ and $\Delta m_e, \Delta \alpha \sim a^{-3}$ deep in the oscillatory regime.
Thus, we can safely ignore any additional effects of rapid oscillations on recombination due to the diminishing amplitude.
For computation purposes, we use an effective fluid description to describe the highly oscillatory regime, which we discuss in the next section.

The other effect of the VFC is the $Y_{\rm He}$ modification, which affects the CMB temperature and polarization power spectra, as shown in Figs.~\ref{fig:Cl_high_mass} and \ref{fig:Cl_low_mass}.
We implement these changes as a correction to the standard BBN prediction used in \texttt{CLASS}:
\begin{equation}
    \label{eq:class_bbn_mod}
    Y_{\rm He} = Y_{\rm He}^{\rm BBN(0)}\left(1 + \dfrac{\Delta Y_{\rm He}}{Y_{\rm He}}\right)\;,
\end{equation}
where $Y_{\rm He}^{\rm BBN(0)}$ is the standard computation (using precomputed interpolation table) in \texttt{CLASS}, which depends on $N_{\rm eff}$ and $\omega_b$.
We calculate ${\Delta Y_{\rm He}}/{Y_{\rm He}}$ using Eq.~\eqref{eq:DY_p}, which depends on $d_i^{(2)}, f_\phi$, and $m_\phi$.

\subsection{$\phi$ Evolution: Field vs Fluid}
The $\phi$ background and perturbation solutions in \texttt{scalarCLASS} are largely based on \texttt{AxiCLASS}.
We briefly review the implementation here, highlighting the changes we made for our purposes.

\subsubsection{Background evolution}
The ``field'' regime marks the period before the onset of oscillation, which is denoted by the time when $H(a) \geq  3m_{\rm eff}$~\cite{Marsh:2015xka}.
During this period, we solve the full Klein-Gordon (KG) equation of the scalar field:
\begin{equation}
\label{eq:eom}
  \ddot{\phi} + 2aH\dot{\phi} + a^2 m_{\rm  eff}^2 \phi = 0\;,
\end{equation}
and we allow for the possibility of setting the contribution from the thermal mass to zero.
The energy density, pressure, and equation of state for the field are
\begin{align}
    \rho_\phi &= {1 \over 2a^2}\dot{\phi}^2 + {1 \over 2}m_{\rm eff}^2\phi^2, \label{eq:rho_phi}\\
    P_\phi &= {1 \over 2a^2}\dot{\phi}^2 - {1 \over 2}m_{\rm eff}^2\phi^2, \label{eq:p_phi}\\
    w_\phi &\equiv {P_\phi \over \rho_\phi}\;,
\end{align}
respectively.

We assume that the abundance of $\phi$ is generated via the misalignment mechanism.
In \texttt{scalarCLASS}, we use a ``shooting mechanism'' to precisely determine the initial $\phi$ to produce a given $\phi$ relic density $\Omega_\phi$.
The evolution of $\phi$ is highly oscillatory for $m_{\rm eff} \gg H$ and is thus numerically expensive to solve.
In the regime $m_\phi > m_{\rm th} \gg H$, the field settles into a matter-like evolution when averaged over those fast oscillations, yielding
\begin{align}
    \label{eq:fluid_eos}
    {1 \over 2}m_{\phi}^2\langle\phi^2\rangle &= {1 \over 2a^2}\langle\dot{\phi}^2\rangle,\\
    \label{eq:fld_rho}
    \rho_\phi &= m_{\rm \phi}^2\langle\phi^2\rangle,\\
    P_\phi &= w_\phi = 0.
\end{align}

\underline{\it Neglecting the thermal mass:}
First, we described our numerical strategy, ignoring the thermal mass.
To determine the evolution of energy density and pressure, \texttt{AxiCLASS} solves Eq.~\eqref{eq:eom} until the time when $H(a_{\rm tr}) = xm_\phi$, with $x = 3$ (by default).
After this transition at scale factor $a_{\rm tr}$ (with $x=3$), $\phi$ sets in on a matter-like evolution~\cite{Marsh:2010wq}.
The energy density is
\begin{equation}
    \label{eq:energy_fluid}
    \rho_\phi(a) = \rho_\phi(a_{\rm tr}) \left(a_{\rm tr} \over a
    \right)^3\;.
\end{equation}
The period after the transition $(a > a_{\rm tr})$ is the ``fluid regime.''
In this regime, the average field value scales as 
$\langle\phi(a)\rangle \sim a^{-3/2}$.

In this work, we study the \emph{oscillatory} evolution of the $\phi$ during BBN and recombination.
Therefore, the treatment of field-to-fluid transition in \texttt{AxiCLASS} is inadequate, since the $\phi$ solution in the fluid regime does not exhibit any oscillation.
In \texttt{scalarCLASS} code, we use a hybrid approach to solve the oscillatory evolution.
We continue solving the $\phi$ equation of motion in Eq.~\eqref{eq:eom} beyond $a_{\rm tr}$ until some later time, denoted by the scale factor $a_{\rm tr2}$, where $m_\phi/H(a_{\rm tr2}) = x_2$, and we choose $x_2 = 100$.
When $a > a_{\rm tr2}$, we take the \emph{average} field value to write, following Eq.~\eqref{eq:fld_rho},
\begin{equation}
  \label{eq:phi_avg}
    \langle \phi \rangle = {\sqrt{\rho_\phi(a_{\rm tr2})}\over m_\phi}\;.
\end{equation}
In this way, we can resolve more than $\mathcal{O}(10)$ oscillations of $\phi$ evolution during $a_{\rm tr} < a < a_{tr2}$, and afterwards, it scales uniformly $\phi \sim a^{-3/2}$.
We find that $x_2 =100$ is sufficient for our purposes of resolving $\mathcal{O}(10)$ oscillations.
Due to the decaying amplitude, the amount of associated VFC is insignificant after $m_{\phi}/H > 100$
for all practical purposes.
Note that for $a_{\rm tr} < a < a_{tr2}$, we use the $\phi$ solution only to determine the amount of VFC.
The evolution of $\rho_\phi$ (and its perturbations) is governed by Eq.~\eqref{eq:energy_fluid}.

In this work, we do not employ the effective fluid approximations that provide a more accurate matching between the field and fluid description~\cite{Passaglia:2022bcr,Liu:2024yne,Cookmeyer:2019rna,Baryakhtar:2024rky,Urena-Lopez:2015gur,Cedeno:2017sou,Urena-Lopez:2023ngt,Moss:2025ymr}.
Many effective fluid approximations produce accurate matching between these different solutions without resolving a single oscillation of the $\phi$ field, which saves computation time.
This, however, is not our primary goal, since we want to track the oscillations of the $\phi$ solutions.
Additionally, the presence of the temperature-dependent thermal mass of the scalar field may non-trivially affect the nature of the effective fluid solutions.

\underline{\it In the presence of the thermal mass:}
In the following, we discuss our strategies for solving $\phi$ in the presence of the thermal mass.
The presence of the thermal mass modifies the scaling of the field $\phi \sim a^{-3/2}$ after oscillation.
When the thermal mass dominates the evolution, $H(a) < m_\phi < m_\mathrm{th}(a)$, the field redshifts faster than matter: $\phi \sim m_{\rm eff}^{-1/2} a^{-3/2}$ after averaging over the oscillations~\cite{Bouley:2022eer,Sibiryakov:2020eir}.
The different temperature dependence of $m_{\rm eff}$ for different couplings $(\alpha,m_e)$ makes $\phi$ scale differently in the thermal-mass-dominated region.
In this work, we do not use any effective description to calculate the field behavior in the thermal-mass-dominated region.
Rather, we solve the Klein-Gordon equations until the effects of the thermal mass become negligible, and the evolution becomes $m_\phi$-dominated.
We choose $x = 100$, so we solve Klein-Gordon equations until $H \geq 100 m_{\rm eff}$ to make sure the field becomes $m_\phi$-dominated at the end, i.e, $m_{\rm th}(a_{\rm tr}) < m_{\rm \phi}$.
We verify that $x = 100$ works well for most of the interesting parameter space relevant for determining our bounds on VFCs.

Fig.~\ref{fig:thmass} depicts the $\phi$ solution in the presence of the thermal mass.
When $m_{\rm th}$ dominates the field evolution [between $H(a)<m_{\rm th}(a)$ and $m_{\rm th}(a)<m_{\phi}$], the envelope of $\phi$ scales much faster than $a^{-3/2}$.
Thus, the temperature dependence of the thermal mass changes the energy density evolution of the scalar field, which needs to be taken into account in an effective fluid description.
The energy density scaling depends on the temperature dependence of the effective mass, which varies with the type of coupling.
The effects of the thermal mass on the effective fluid description merit a dedicated study and can have wide-ranging applications for models of ULDM coupling with SM.

\subsubsection{Perturbation evolution}

For the high-mass analysis, it is sufficient to treat the ULDM perturbations as CDM-like for our CMB analyses.
We only compute the background $\phi$ evolution to determine the amount of $\delta Y_{\rm He}$, which sources the only relevant effect of the VFC in the CMB power spectra.

We do compute the ULDM perturbations in the context of the low-mass analysis and neglect the effects of the thermal mass, which we discuss further in Appendix.~\ref{app:mcmctherm}.
Our implementation of the perturbations is based on \texttt{AxiCLASS}~\cite{Poulin:2018dzj,Smith:2019ihp}.
Linear perturbation equations of the ULDM field can be derived using the perturbed Klein-Gordon equation~\cite{Hu:1998kj}.
The alternative approach, which \texttt{AxiCLASS} employs, is to solve the equations for the density fluctuations $\delta_\phi$ and velocity divergences $\theta_\phi$ of the scalar field, treating it as a fluid for the entirety of the evolution~\cite{Hu:1998kj}.
The linear perturbation equations in synchronous gauge are given by
\begin{align}
    \dot{\delta}_\phi &= - (1+w_\phi)\left(\theta_\phi + {\dot{h} \over 2}\right) - 3(c_s^2 - w_\phi)\mathcal{H}\delta_\phi - 9(1+w_\phi)(c_s^2 - c_\phi^2)\mathcal{H}{\theta_\phi \over k^2}\;,\\
    \dot{\theta}_\phi &= - (1 - 3c_s^2)\mathcal{H}\theta_\phi + {c_s^2k^2 \over 1+ w_\phi}\delta_\phi \;,
\end{align}
where $\mathcal{H} = aH$ is the comoving Hubble rate and
\begin{equation}
    c_\phi^2 \equiv \frac{\dot{P}_\phi}{\dot{\rho}_\phi} = w_\phi - \frac{\dot{w}_\phi}{3(1+w_\phi)\mathcal{H}}. 
\end{equation}
is the adiabatic sound speed of $\phi$, since the energy density scales as $\rho_\phi \sim a^{-3(1+w_\phi)}$.
Using Eqs.~\eqref{eq:rho_phi} and \eqref{eq:p_phi}, along with the Eq.~\ref{eq:eom}, the adiabatic sound speed can be written as
\begin{equation}
    \label{eq:cad2}
    c_\phi^2 \equiv \dfrac{\dot{P}_\phi}{\dot{\rho}_\phi} = 1 + \dfrac{2m_\phi^2\phi}{3H\dot{\phi}/a}\;.
\end{equation}
In \texttt{scalarCLASS}, we use the solution for the $\phi$ evolution to compute $c_\phi^2$ from Eq.~\eqref{eq:cad2}, which we then use to compute $w_\phi$ for better numerical stability.
After the onset of oscillation, $w_\phi = c_\phi^2 = 0$ as the field locks into a matter-like evolution.
The effective sound speed $c_s^2$ is given by~\cite{Hu:1998kj,Marsh:2010wq,Hlozek:2014lca,Poulin:2018dzj}
\begin{equation}
    \label{eq:cs2}
    c_s^2 \equiv \frac{\delta P_\phi}{\delta\rho_\phi} = {k^2 \over k^2 + 4a^2m_\phi^2}\;.
\end{equation}

\section{Impact of thermal effects on the low-mass analysis}

\label{app:mcmctherm}
\begin{figure}
    \centering
    \includegraphics[width=0.85\linewidth]{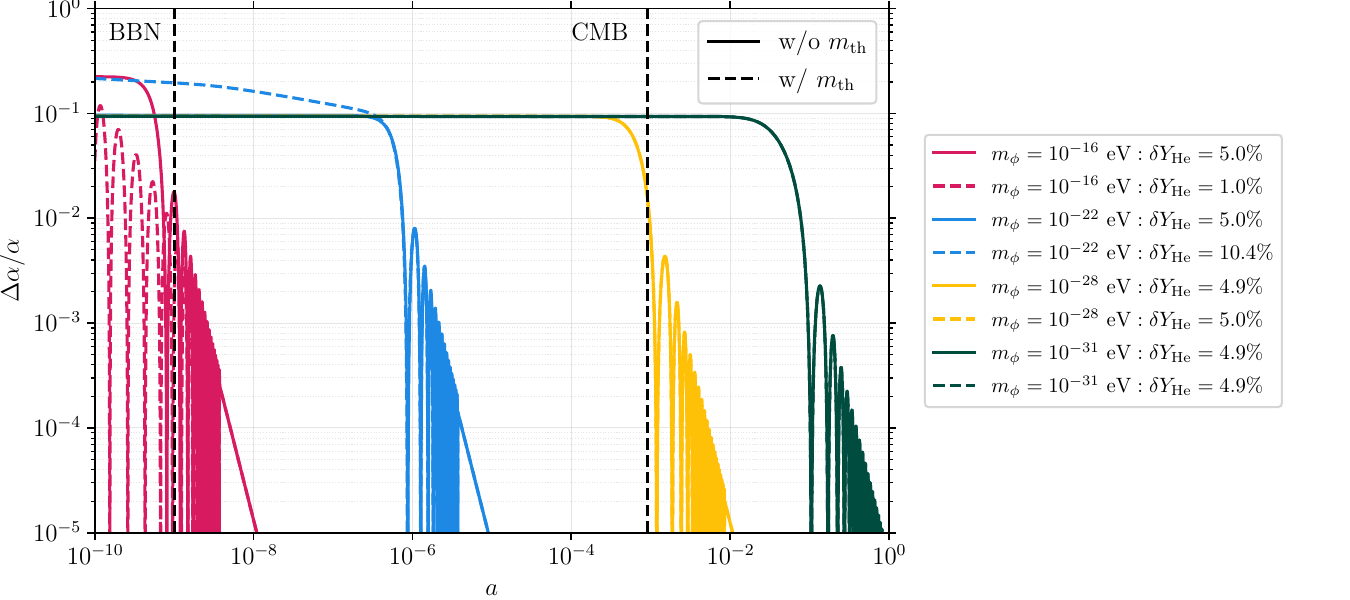}
    \caption{Evolution of the variation of $\alpha$ as a function of scale factor for different masses of $\phi$, both with (dashed) and without (solid) the contribution of the thermal mass.
      As in Fig.~\ref{fig:variation_w_diff_mass}, we choose the amount of variation such that $\delta Y_\mathrm{He} = 5\%$ for the case without $m_\mathrm{th}$.
      We fix $f_\phi = 1$ when computing the thermal-mass effects and provide the corresponding $\delta Y_{\rm He}$ for each mass in the legend.
    The thermal mass modifies the field evolution at early times and thus affects the VFC mostly during BBN.
    Note that for $m_\phi = 10^{-16}$ eV, including the thermal effects results in a lower $Y_{\rm He}$, while the opposite happens for $m_\phi = 10^{-22}$ eV.
    For even lighter masses, thermal effects have a negligible impact on the VFC during BBN.
    For all masses, the VFC during recombination is unaffected by thermal effects.
    }
    \label{fig:variation_w_different_mass_therm}
\end{figure}

As we discuss at the end of Sec.~\ref{sec:cmb-combined}, we neglect the thermal mass of the scalar field for the analysis in the low-mass region in Sec.~\ref{sec:mcmc-low}.
In this appendix, we discuss the validity of this approximation.

The thermal mass affects the evolution of $\phi$ more strongly at early times (as shown in Fig.~\ref{fig:thmass}) due to its strong temperature dependence.
The contribution of the thermal mass fades over time, allowing $m_\phi$ alone to dictate the evolution at later times.
In Fig.~\ref{fig:variation_w_different_mass_therm}, we show the evolution of the variation of $\alpha$ both with (dashed) and without (solid; identical to Fig.~\ref{fig:variation_w_diff_mass}) the effects of the thermal mass.
We fix $f_\phi = 1$ and choose the coupling $d_\alpha^{(2)}$ such that $\delta Y_\mathrm{He} \approx 5\%$ when ignoring the thermal mass.
The thermal mass is important for the heavier masses, as the field oscillations start at earlier times, relevant for BBN.
Therefore, accounting for $m_{\rm th}$ modifies the $Y_\mathrm{He}$ prediction for the larger values of $m_\phi$.
However, for smaller values of $m_\phi$, the effects of $m_{\rm th}$ at BBN are small, and the effects at recombination are negligible.

The importance of the thermal mass for the various values of $m_\phi$ in Fig.~\ref{fig:variation_w_different_mass_therm} may seem counterintuitive in light of Fig.~\ref{fig:thmass}: for smaller $m_\phi$, the field is under influence of $m_{\rm th}$ longer until the condition $m_{\rm th}(a) < m_\phi$ is met, but this situation holds for a fixed coupling.
Recall that the amount of variation depends on the combination $f_\phi d_i^{(2)}$, while the thermal mass depends only on the coupling $d_i^{(2)}$.

During early evolution $[m_{\rm th}(a) > H(a) > m_\phi ]$, when the field is under the influence of $m_{\rm th}$, the amplitude of the field decreases rapidly compared to the case in which the thermal mass is neglected.
Thus, for a fixed initial field value $\phi_i$, including the effects of $m_{\rm th}$ decreases the abundance at late times.
Alternatively, achieving a set abundance at late times requires a larger $\phi_i$, increasing the VFC during BBN.
Since we fix $f_\phi = 1$ for Fig.~\ref{fig:variation_w_different_mass_therm}, the value of the coupling $d_i^{(2)}$ is limited and thus the impact of the thermal mass is limited.

\begin{figure}
    \centering
    \includegraphics[width=0.49\linewidth]{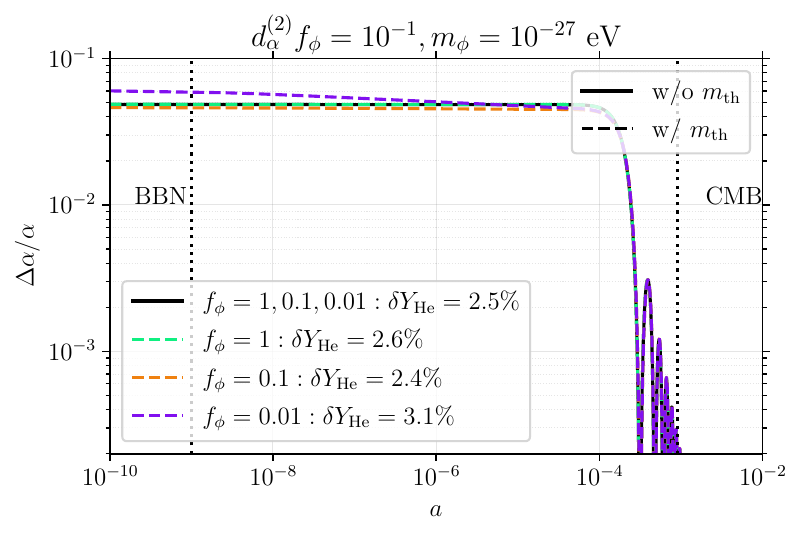}
    \includegraphics[width=0.49\linewidth]{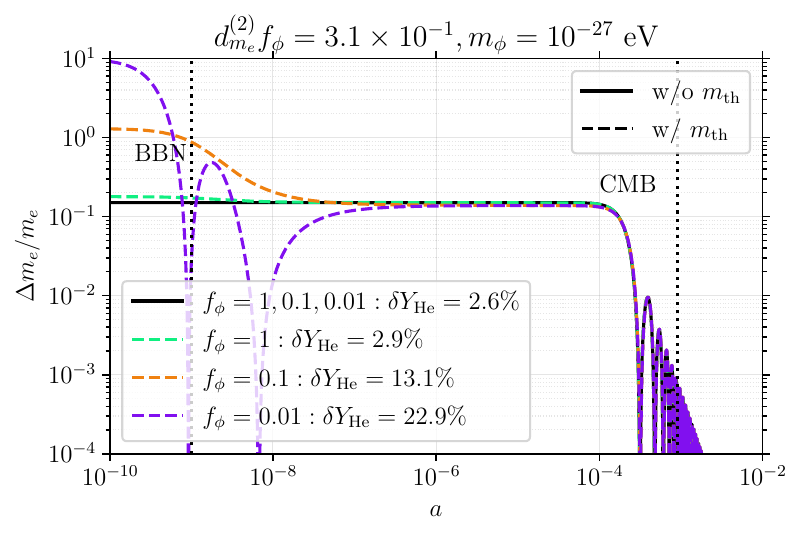}
    \caption{ Effect of $m_{\rm th}$ on VFC  for different ULDM fraction $f_\phi$. Variation in the presence of $m_{\rm th}$ is shown in dashed lines, while the solid line shows the VFC for bare mass alone. For all the curves, we have kept $d_i^{(2)} f_\phi$ fixed while varying $f_\phi$. Fixed $d_i^{(2)} f_\phi$ gives an identical amount of variations for each $f_\phi$ for the bare mass case, since $\Delta m_e, \Delta\alpha \propto d_i^{(2)}f_\phi$. However, $m_{\rm th}$ only depends on $d_i^{(2)}$ and smaller $f_\phi$ requires a larger $d_i^{(2)}$ for our choice of fixed $d_i^{(2)} f_\phi$. Thus, for smaller $f_\phi$ the effects of thermal mass are larger, as can be seen from the dashed lines. $m_{\rm th}$ influences the field evolution at early time, thus the prediction for $\delta Y_{\rm He}$ gets modified. However, even for small $f_\phi$, the thermal mass effects on VFC during recombination are negligible. 
    }
    \label{fig:thmass_small_m}
\end{figure}

The effects of $m_\mathrm{th}$ become more important when $f_\phi$ is reduced.
Fig.~\ref{fig:thmass_small_m} shows the effects of the thermal mass on the evolution of small-mass ULDM for different $f_\phi$ with a fixed $d_i^{(2)}f_\phi$.
Without thermal effects, the modifications of $\alpha$ or $m_e$ are only dependent on coupling times the value of $\phi^2$:
\begin{equation}
    \label{eq:var_scaling}
    \Delta m_e, \Delta\alpha \propto d_i^{(2)}\phi^2 \propto d_i^{(2)} \omega_\phi \propto d_i^{(2)}f_\phi\;,
\end{equation}
where $\omega_\phi$ is the $\phi$ physical energy density. 
Thus, the VFC profile is independent of $f_\phi$ for a fixed $d_i^{(2)}f_\phi$. Introducing $m_{\rm th}$ breaks the degeneracy between $d_i^{(2)}$ and $f_\phi$.
For a fixed $d_i^{(2)} f_\phi $, a smaller $f_\phi$ necessitates a larger coupling $d_i^{(2)}$, which results in an enhanced effect from the thermal mass. However, in all the cases depicted in Fig.~\ref{fig:thmass_small_m}, the introduction of $m_{\rm th}$ only affects the field evolution at higher redshift, affecting the BBN computation of $Y_\mathrm{He}$. The late-time behaviors, especially during recombination, are identical to the no-thermal-mass case. More generally, the field evolution down to $f_\phi\sim 10^{-3}$, in the parameter space of interest, is unaffected by $m_{\rm th}$. Thus, incorporating $m_{\rm th}$ for the low mass case does not affect the constraints due to modifications of recombination. Since the novelty in this mass range is modified recombination, we can safely drop $m_{\rm th}$ when deriving the constraint from the low mass MCMC analysis.

We end this section with a comment about the growth of structure.
In the presence of the thermal mass, the $\phi$ perturbation equations are coupled with the perturbations of the photon bath.
This changes the evolution of the $\phi$ perturbations at small scales, which enter the horizon at higher redshift when the thermal mass dominates the evolution.
These effects may have interesting implications for the small-scale ULDM power spectrum and can modify the growth of structure for scales where thermal effects are important.
These modifications are expected to be small for the scales relevant to CMB.
Further, the effects of the energy transfer from bath to ULDM are also small as explained in the main text.
Thus, ULDM perturbation evolution for the low-mass region analysis remains unchanged from the standard evolution.

\section{Triangle plots and additional tables}
\label{app:traingle_plots}

\subsection{High-mass analysis}
\begin{figure}
    \centering
    \includegraphics[width=0.8\linewidth]{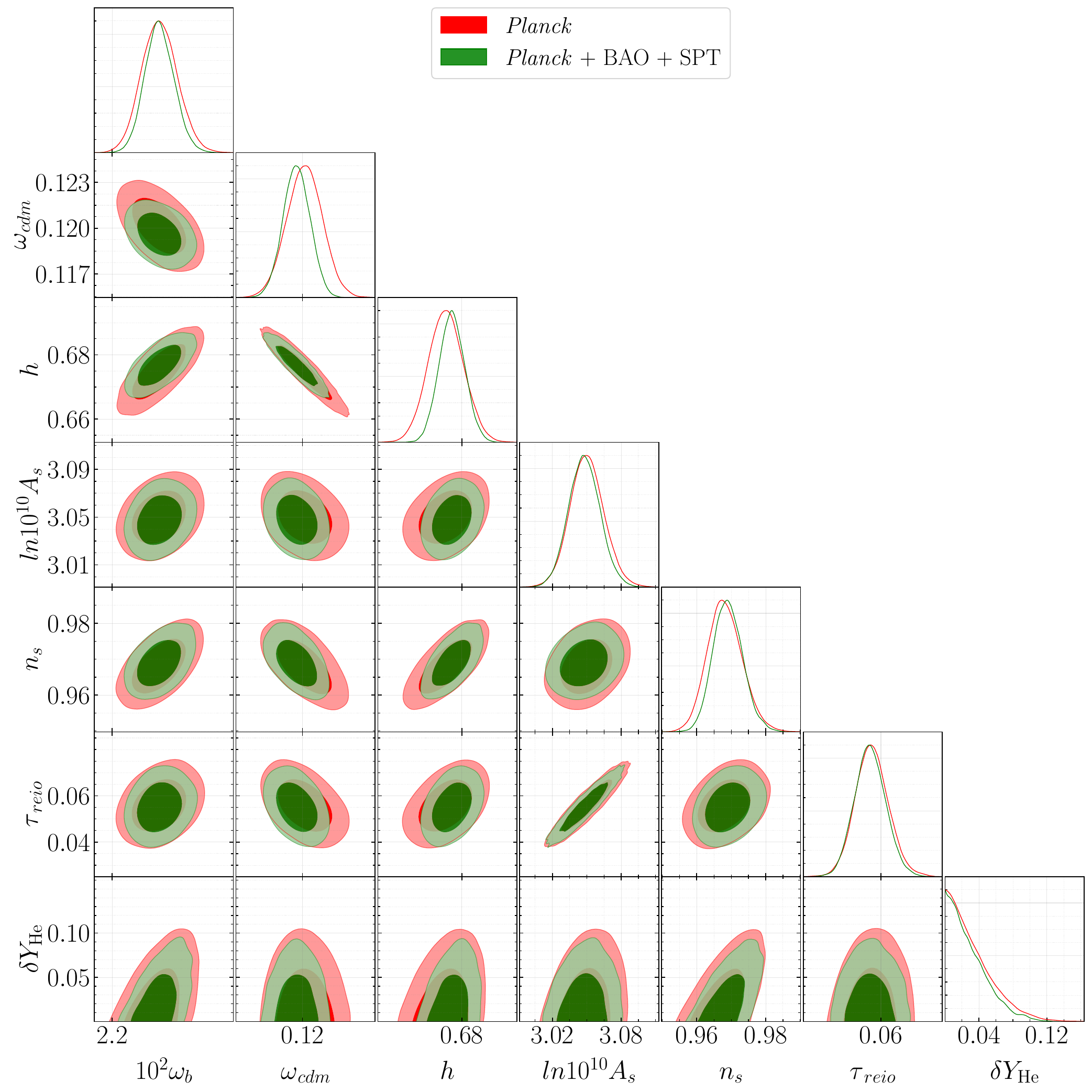}
    \caption{
    Marginalized $1\sigma$ and $2\sigma$ posteriors for the parameters of the $\Lambda$CDM + $\delta Y_{\rm He}$ MCMC analysis relevant for VFC induced by the high-mass ULDM.}
    \label{fig:full_try_dyhe}
\end{figure}
The triangle plot for all the relevant parameters for the MCMC analysis, where we varied $\delta Y_{\rm He}$ along with $\Lambda$CDM parameters, is shown in Fig.~\ref{fig:full_try_dyhe}.

\subsection{Low-mass analysis} 
\begin{figure}
    \centering
    \includegraphics[width=\linewidth]{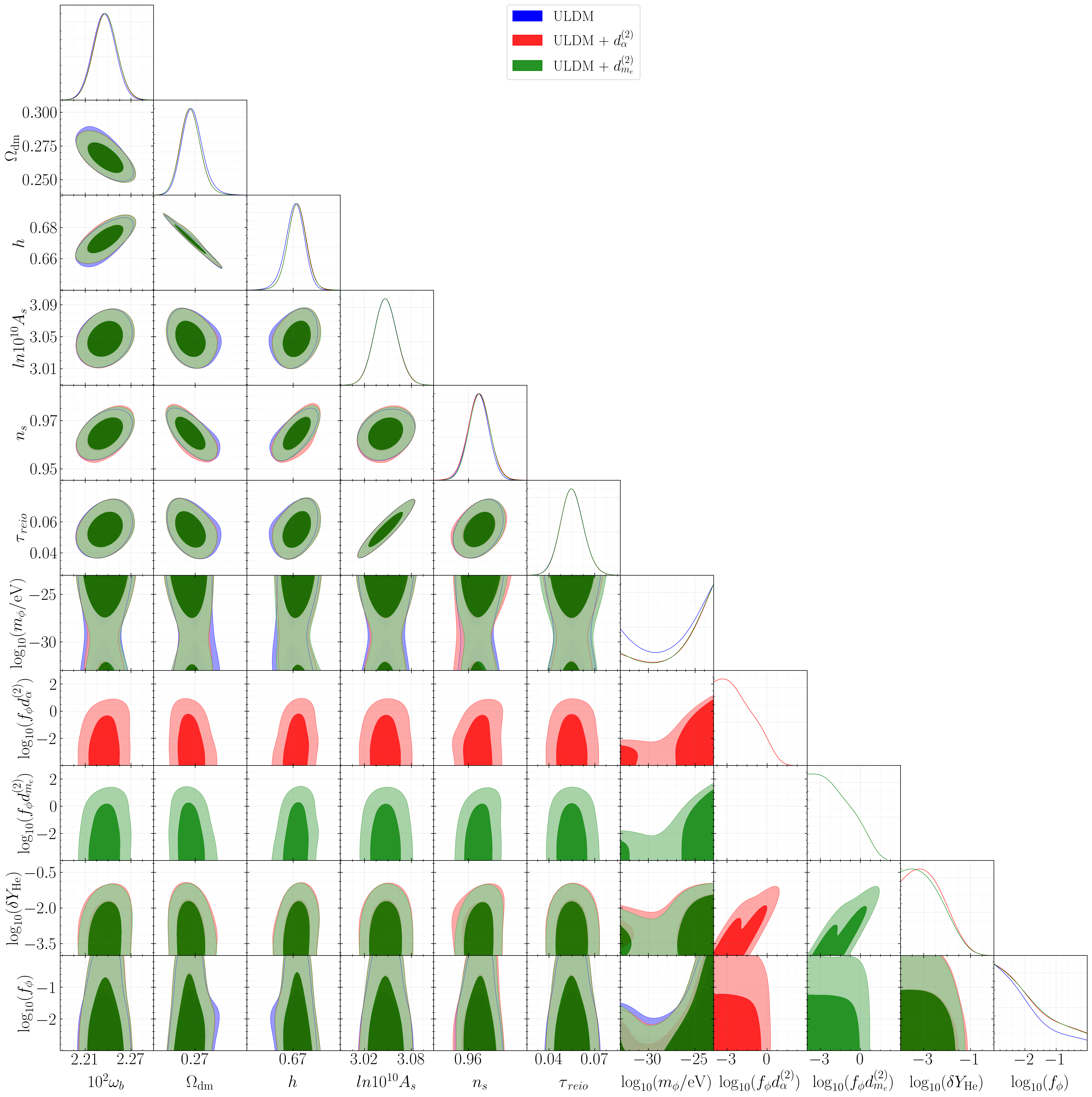}
    \caption{
    $68\%$ and $98\%$ C.L.~posteriors for the cosmological parameters for the low-mass analyses of the pure ULDM, ULDM + $\alpha$ and ULDM + $m_e$ cases using the \textit{Planck} dataset.  }
    \label{fig:ftri_planck}
\end{figure}

\begin{figure}
    \centering
    \includegraphics[width=\linewidth]{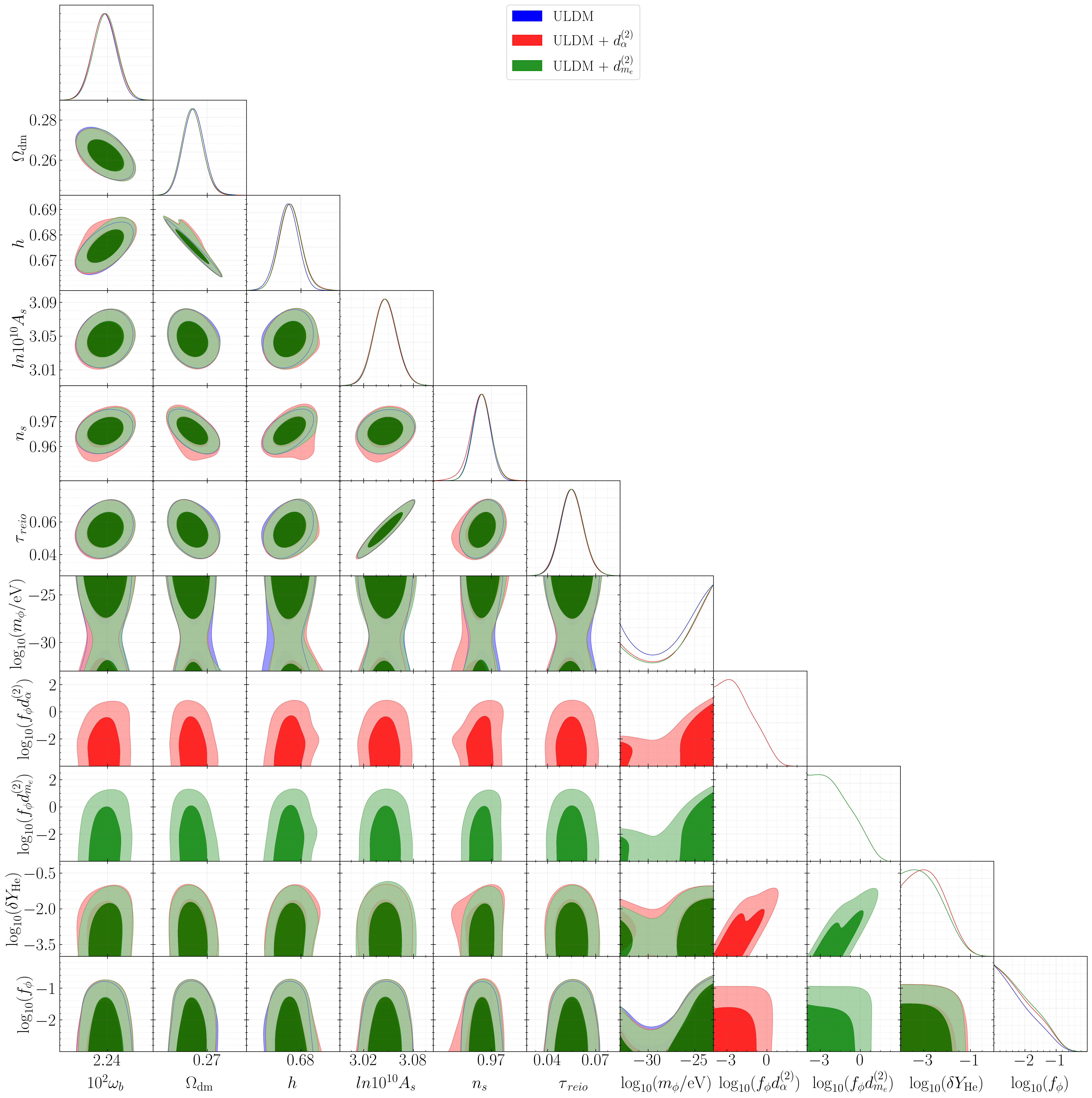}
    \caption{
    $68\%$ and $98\%$ C.L.~posteriors for the cosmological parameters for the low-mass analyses of the pure ULDM, ULDM + $\alpha$ and ULDM + $m_e$ cases using the \textit{Planck} + BAO + SPT datasets.  }
    \label{fig:ftri_psb}
\end{figure}

\begin{figure}
    \centering
    \includegraphics[width=\linewidth]{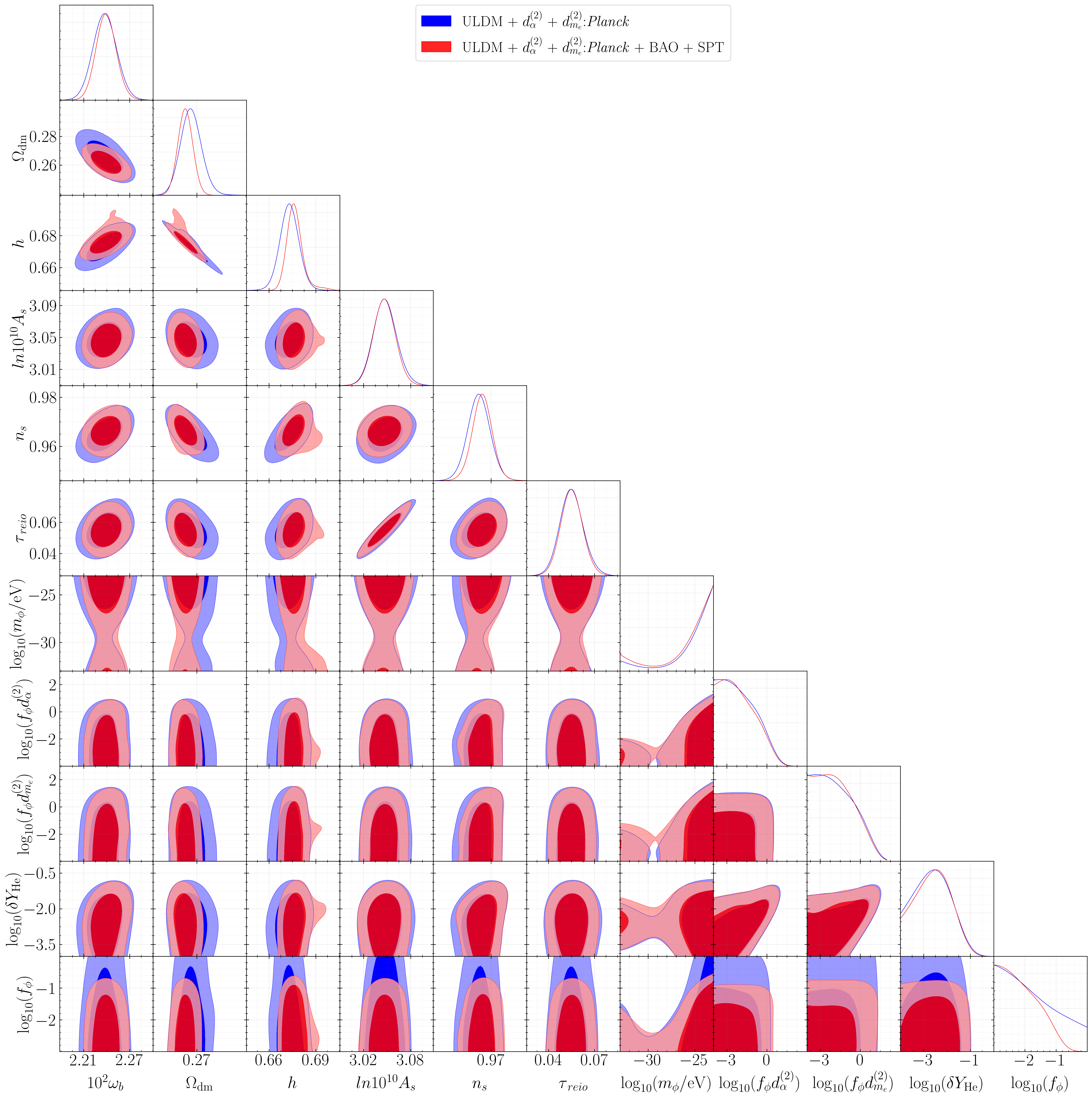}
    \caption{
    $68\%$ and $98\%$ C.L.~posteriors for the cosmological parameters for low-mass analysis of the simultaneous variation $\alpha$ and $m_e$ for the \textit{Planck} and the \textit{Planck} + BAO + SPT datasets.  }
    \label{fig:ftri_comb}
\end{figure}

For the low-mass analysis, constraints for all relevant cosmological parameters are provided in Tables~\ref{tab:p} and \ref{tab:psb}. They also include the analyses for the simultaneous variation of $d_\alpha^{(2)}$ and $d_{m_e}^{(2)}$.
We show the triangle plots for all the cases for \textit{Planck} and \textit{Planck} + BAO + SPT dataset in Fig.~\ref{fig:ftri_planck} and Fig.~\ref{fig:ftri_psb}, respectively. The triangle plot for the simultaneous variation of $d_\alpha^{(2)}$ and  $d_{m_e}^{(2)}$ is shown in Fig.~\ref{fig:ftri_comb}. 

\renewcommand{\arraystretch}{1.1}
\begin{table}[]
    \centering
    \begin{tabular}{|c|c|c|c|c|}
    \hline
         & \multicolumn{4}{c|}{\textit{Planck}} \\
         \hline
         & ULDM & ULDM + $d_\alpha^{(2)}$ & ULDM + $d_{m_e}^{(2)}$ & ULDM + $d_\alpha^{(2)}$ + $d_{m_e}^{(2)}$ \\
         \hline
         $10^2 \omega_{b}$ & $ 2.23\pm 0.01$ & $ 2.24\pm 0.02$ & $ 2.24\pm 0.02$ & $ 2.24\pm 0.02$\\
\hline
$\Omega_{\rm DM}$ & $ 0.268^{+0.007}_{-0.008}$ & $ 0.267^{+0.007}_{-0.008}$ & $ 0.267^{+0.007}_{-0.008}$ & $ 0.266^{+0.007}_{-0.008}$\\
\hline
$h$ & $ 0.671^{+0.006}_{-0.005}$ & $ 0.673\pm 0.006$ & $ 0.672\pm 0.006$ & $ 0.673\pm 0.006$\\
\hline
$\ln 10^{10}A_{s }$ & $ 3.05\pm 0.01$ & $ 3.05\pm 0.01$ & $ 3.05\pm 0.01$ & $ 3.05^{+0.01}_{-0.02}$\\
\hline
$n_{s }$ & $ 0.964\pm 0.004$ & $ 0.964\pm 0.005$ & $ 0.965\pm 0.004$ & $ 0.965\pm 0.005$\\
\hline
$\tau{}_{\rm reio}$ & $ 0.055\pm 0.008$ & $ 0.055\pm 0.008$ & $ 0.055\pm 0.008$ & $ 0.055^{+0.007}_{-0.008}$\\
\hline
$\log_{10}(m_\phi /{\rm eV})$ & $-$ & $> -26.0$ & $> -26.0$ & $> -25.2$\\
\hline
$\log_{10}(f_{\phi}d_\alpha^{(2)})$ & $-$ & $< -1.57$ & $-$ & $< -1.39$\\
\hline
$\log_{10}(f_\phi d_{m_e}^{(2)})$ & $-$ & $-$ & $< -1.28$ & $< -1.11$\\
\hline
$\log_{10}(f_{\phi})$ & $< -1.83$ & $< -1.63$ & $< -1.61$ & $< -1.45$\\
\hline
\hline
$\log_{10}(\delta Y_{\rm He})$ & $-$ & $ < -2.11 $ & $ <-2.29$ & $ <-1.69$\\
\hline
$\sigma_8$ & $ 0.806^{+0.01}_{-0.006}$ & $ 0.807^{+0.01}_{-0.006}$ & $< 0.812$ & $ 0.808^{+0.01}_{-0.005}$\\
\hline
    \end{tabular}
    \caption{ $1\sigma$ measurements and $68\%$ C.L. constraints on the parameters for the low-mass MCMC analysis with the \textit{Planck} dataset.}
    \label{tab:p}
\end{table}
\begin{table}[]
    \centering
    \begin{tabular}{|c|c|c|c|c|}
    \hline
         & \multicolumn{4}{c|}{\textit{Planck} + BAO + SPT} \\
         \hline
         & ULDM & ULDM + $d_\alpha^{(2)}$ & ULDM + $d_{m_e}^{(2)}$ & ULDM + $d_\alpha^{(2)}$ + $d_{m_e}^{(2)}$ \\
         \hline
$10^2 \omega_{b}$ & $ 2.24\pm 0.01$ & $ 2.24\pm 0.01$ & $ 2.24\pm 0.01$ & $ 2.24\pm 0.01$\\
\hline
$\Omega_{\rm DM}$ & $ 0.263\pm 0.005$ & $ 0.262\pm 0.005$ & $ 0.262\pm 0.005$ & $ 0.262\pm 0.005$\\
\hline
$h$ & $ 0.675\pm 0.004$ & $ 0.676\pm 0.004$ & $ 0.676^{+0.004}_{-0.005}$ & $ 0.677^{+0.004}_{-0.005}$\\
\hline
$\ln 10^{10}A_{s }$ & $ 3.05\pm 0.01$ & $ 3.05\pm 0.01$ & $ 3.05\pm 0.01$ & $ 3.05\pm 0.01$\\
\hline
$n_{s }$ & $ 0.966\pm 0.004$ & $ 0.966\pm 0.004$ & $ 0.966\pm 0.004$ & $ 0.966\pm 0.004$\\
\hline
$\tau{}_{\rm reio}$ & $ 0.055\pm 0.007$ & $ 0.055\pm 0.007$ & $ 0.055\pm 0.007$ & $ 0.055\pm 0.007$\\
\hline
$\log_{10}(m_\phi /{\rm eV})$ & $-$ & $> -26.4$ & $> -26.3$ & $> -25.6$\\
\hline
$\log_{10}(f_{\phi}d_\alpha^{(2)})$ & $-$ & $< -1.66$ & $-$ & $< -1.49$\\
\hline
$\log_{10}(f_\phi d_{m_e}^{(2)})$ & $-$ & $-$ & $< -1.37$ & $< -1.21$\\
\hline
$\log_{10}(f_{\phi})$ & $< -2.01$ & $< -1.92$ & $< -1.91$ & $< -1.83$\\
\hline
\hline
$\log_{10}(\delta Y_{\rm He})$ & $-$ & $ < -2.03 $ & $ <-2.27$ & $ <-1.66$\\
\hline
$\sigma_8$ & $ 0.805^{+0.009}_{-0.006}$ & $ 0.806^{+0.009}_{-0.006}$ & $ 0.807^{+0.009}_{-0.006}$ & $ 0.808^{+0.008}_{-0.006}$\\
\hline
    \end{tabular}
    \caption{$1 \sigma$ measurements and $68\%$ C.L. constraints on the parameters for the low-mass MCMC analysis with the \textit{Planck} + BAO + SPT dataset.}
    \label{tab:psb}
\end{table}

\bibliography{references.bib}

\end{document}